\documentclass[preprint,epsf,showpacs]{revtex4}
\usepackage{amsmath}
\usepackage{epsfig}
\usepackage{epstopdf}

\begin{document}

\title{Dynamic phase transitions in the presence of quenched randomness}

\author{Erol Vatansever$^1$}

\author{Nikolaos~G. Fytas$^2$}

\affiliation{$^1$Department of Physics, Dokuz Eyl\"{u}l
University, TR-35160, Izmir-Turkey}

\affiliation{$^2$Applied Mathematics Research Centre, Coventry
University, Coventry  CV1 5FB, United Kingdom}

\date{\today}

\begin{abstract}
We present an extensive study of the effects of quenched disorder
on the dynamic phase transitions of kinetic spin models in two
dimensions. We undertake a numerical experiment performing
Monte Carlo simulations of the square-lattice
random-bond Ising and Blume-Capel models under a periodically
oscillating magnetic field. For the case of the Blume-Capel model
we analyze the universality principles of the dynamic
disordered-induced continuous transition at the low-temperature
regime of the phase diagram. A detailed finite-size scaling
analysis indicates that both nonequilibrium phase transitions
belong to the universality class of the corresponding equilibrium
random Ising model.
\end{abstract}

 \maketitle

\section{Introduction}
\label{sec:introduction}

In the last sixty years our understanding of equilibrium critical
phenomena has developed to a point where well-established results
are available for a wide variety of systems. In particular, the
origin and/or the difference between equilibrium universality
classes is by now well understood. This observation also partially
holds for systems under the presence of quenched disorder.
However, far less is known for the physical mechanisms underlying
the nonequilibrium phase transitions of many-body interacting
systems that are far from equilibrium and clearly a general
classification of nonequilibrium phase transitions into
nonequilibrium universality classes is missing.

We know today that when a ferromagnetic system, below its Curie
temperature, is exposed to a time-dependent oscillating magnetic
field, it may exhibit a fascinating dynamic magnetic
behavior~\cite{Tome}. In a typical ferromagnetic system being
subjected to an oscillating magnetic field, there occurs a
competition between the time scales of the applied-field period
and the metastable lifetime, $\tau$, of the system. When the
period of the external field is selected to be smaller than
$\tau$, the time-dependent magnetization tends to oscillate around
a nonzero value, which corresponds to the dynamically ordered
phase. In this region, the time-dependent magnetization is not
capable of following the external field instantaneously. However,
for larger values of the period of the external field, the system
is given enough time to follow the external field, and in this
case the time-dependent magnetization oscillates around its zero
value, indicating a dynamically disordered phase. When the period
of the external field becomes comparable to $\tau$, a dynamic
phase transition takes place between the dynamically ordered and
disordered phases.

Throughout the years, there have been several
theoretical~\cite{Lo, Zimmer, Acharyya1, Chakrabarti, Acharyya2,
Acharyya3, Buendia1, Buendia2, Fujisaka, Jang1, Jang2, Shi, Punya,
Riego, Keskin1, Keskin2, Robb1,Deviren, Yuksel1, Yuksel2,
Vatansever1} and experimental
studies~\cite{He,Robb,Suen,Berger,Riego1} dealing with dynamic
phase transitions, as well as with the hysteresis properties of
magnetic materials. The main conclusion emerging is that both the
amplitude and the period of the time-dependent magnetic field play
a key role in dynamical critical phenomena (in addition to the
usual temperature parameter). Furthermore, the characterization of
universality classes in spin models driven by a time-dependent
oscillating magnetic field has also attracted a lot of interest
lately~\cite{Sides1,Sides2,Korniss,Buendia3,
Park,Park2,Tauscher,Buendia4,Vatansever2,Vatansever_Fytas}. Some
of the main results are listed below:

\begin{itemize}

\item The critical exponents of the kinetic Ising model were found
to be compatible to those of the equilibrium Ising model at both
two- (2D) and three dimensions
(3D)~\cite{Sides1,Sides2,Korniss,Vatansever2,Park}.

\item Buend\'{i}a and Rikvold using soft Glauber dynamics
estimated the critical exponents of the 2D Ising model and
provided strong evidence that the characteristics of the dynamic
phase transition are universal with respect to the choice of the
stochastic dynamics~\cite{Buendia3}.

\item The role of surfaces at nonequilibrium phase transitions in
Ising models has been elucidated by Park and Pleimling: The
nonequilibrium surface exponents were found to be different than
equilibrium critical surface ones~\cite{Park2}.

\item Experimental evidence by Riego \emph{et al.}~\cite{Riego1}
and numerical results by Buend\'{i}a and Rikvold~\cite{Buendia4}
verified that the equivalence of the dynamic phase transition to
an equilibrium phase transition is limited to the area near the
critical period and for zero bias.

\item Numerical simulations by Vatansever and Fytas showed that
the nonequilibrium phase transition of the spin-1 Blume-Capel
model belongs to the universality class of the equilibrium Ising
counterpart (at both 2D and 3D)~\cite{Vatansever_Fytas}. General
and very useful features of the dynamic phase transition of the
Blume-Capel model can also be found in Refs.~\cite{Buendia1,
Keskin1, Keskin2, Deviren, ShiWei, Acharyya4}.

\end{itemize}

The above results in 2D and 3D kinetic Ising and Blume-Capel
models establish a mapping between the universality principles of
the equilibrium and dynamic phase transitions of spin-1/2 and
spin-1 models. They also provide additional support in favor of an
earlier investigation of a Ginzburg-Landau model with a
periodically changing field~\cite{Fujisaka}, as well as with the
symmetry-based arguments of Grinstein \emph{et al.} in
nonequilibrium critical phenomena~\cite{Grinstein}.

Motivated by the current literature, in the present work we
attempt to shed some light on the effect of quenched disorder on
dynamic phase transitions. To the best of our knowledge, with the
exception of a few mean-field and effective-field theory
treatments of the
problem~\cite{Gupinar12,Gupinar12b,Akinci12,Vatansever13,Vatansever13b,Vatansever15},
there exists no dedicated (numerical) work. However,
what we have mainly learned from the previous studies
on the topic is that the dynamic character of a typical magnetic
system driven by a time-dependent magnetic field sensitively
depends on the amount of disorder, accounting for 
reentrant phenomena and dynamic
tricritical points~\cite{Akinci12}. In the current work we use as test-case
platforms for our numerical experiment the Ising and Blume-Capel
models on the square lattice under a time-dependent magnetic
field, diffusing disorder in the ferromagnetic exchange
interactions. For the case of the Blume-Capel model we focus on
the disordered-induced continuous dynamic transition at the
low-temperature regime of the phase diagram. In a nutshell, our
results indicate that the dynamic phase transitions of both the
random-bond Ising and Blume-Capel models belong to the
universality class of the equilibrium random Ising model.

The outline of the remainder parts of the paper is as follows: In
Sec.~\ref{sec:models} we introduce the disordered versions of the
Ising and Blume-Capel models and in
Sec.~\ref{sec:observables} the thermodynamic observables necessary for the
application of the finite-size scaling analysis. The details our simulation
protocol are given in Sec.~\ref{sec:numerics} and the numerical
results and discussion in Sec.~\ref{sec:results}.
Finally, Sec.~\ref{sec:conclusions} contains a summary of our conclusions.

\section{Models}
\label{sec:models}

We consider the square-lattice random-bond Ising and Blume-Capel
(BC) models under a time-dependent oscillating magnetic field, described by the 
following Hamiltonians
\begin{equation}\label{eq:1}
 \mathcal{H}^{\rm (Ising)} = -\sum_{\langle xy \rangle}J_{xy}\sigma_{x}\sigma_{y}-h(t)\sum_{x}\sigma_{x} ,
\end{equation}
and
\begin{equation}\label{eq:2}
 \mathcal{H}^{\rm (BC)} = -\sum_{\langle xy \rangle}J_{xy}\sigma_{x}\sigma_{y}+\Delta \sum_{x}\sigma_{x}^2-h(t)\sum_{x}\sigma_{x} .
\end{equation}
In the above Eqs.~(\ref{eq:1}) and (\ref{eq:2}) $\langle xy \rangle$ indicates summation over
nearest neighbors and the spin variable $\sigma_{x}$ takes on the
values $\{ \pm 1 \}$ for the Ising and $\{-1,0,+1\}$ for the BC
model, respectively. The couplings $J_{xy}>0$ denote the random
ferromagnetic exchange interactions, drawn from a
bimodal distribution of the form
\begin{equation} \label{eq:3}
\mathcal{P}(J_{xy})=\frac{1}{2}~[\delta(J_{xy}-J_{1})+\delta(J_{xy}-J_{2})].
\end{equation}
Following Refs.~\cite{Malakis1,Malakis2,Malakis3}, we choose
$J_{1}+J_{2} = 2$ and $J_{1} > J_{2} > 0$, so that $r = J_{2} /
J_{1}$ defines the disorder strength; for $r = 1$ the pure systems
are recovered. A clear advantage of using the bimodal distribution~(\ref{eq:3}) 
is that the critical temperature $T_{\rm c}$
of the random Ising model is exactly known from duality
relations as a function of the disorder
strength $r$ via~\cite{Fisch,Kinzel}
\begin{equation}
\label{eq:4} \sinh{(2J_{1}/T_{\rm c})}\sinh{(2rJ_{1}/T_{\rm c})} =
1.
\end{equation}
For the case of the Blume-Capel Hamiltonian~(\ref{eq:2}) $\Delta$
denotes the crystal-field coupling that controls the density of
vacancies ($\sigma_{x} = 0$). For $\Delta \rightarrow -\infty$
vacancies are suppressed and the model becomes equivalent to the
Ising model. Finally, the term $h(t)$ corresponds to a spatially
uniform periodically oscillating magnetic field, so that all
lattice sites are exposed to a square-wave magnetic field with
amplitude $h_{0}$ and half period
$t_{1/2}$~\cite{Korniss,Buendia3,Park}.

A brief description of the Blume-Capel model's phase diagram
together with some necessary pinpoints of the current literature
with respect to the effect of disorder on its critical behavior
may be useful here: The phase diagram of the equilibrium pure
Blume-Capel model in the crystal-field -- temperature plane
consists of a boundary that separates the ferromagnetic from the
paramagnetic phase. The ferromagnetic phase is characterized by an
ordered alignment of $\pm 1$ spins. On the other hand, the
paramagnetic phase can be either a completely disordered
arrangement at high temperature or a $\pm1$-spin gas in a $0$-spin
dominated environment for low temperatures and high crystal
fields. At high temperatures and low crystal fields, the
ferromagnetic-paramagnetic transition is a continuous phase
transition in the Ising universality class, whereas at low
temperatures and high crystal fields the transition is of
first-order character~\cite{Capel,Blume}. The model is thus a
classic and paradigmatic example of a system with a tricritical
point $[\Delta_{\rm t},T_{\rm t}]$~\cite{Lawrie}, where the two
segments of the phase boundary meet. A detailed reproduction of
the phase diagram of the model can be found in
Ref.~\cite{Zierenberg} and an accurate estimation of the location
of the tricritical point has been given in Ref.~\cite{Kwak}:
$[\Delta_{\rm t},T_{\rm t}] = [1.9660(1), 0.6080(1)]$. A lot of
work has been also devoted in understanding the effects of
quenched bond randomness on the universality aspects of the
Blume-Capel model, especially in two dimensions, where any
infinitesimal amount of disorder drives the first-order transition
at the low-temperature regime to a continuous transition.
Quantitative phase diagrams of the random-bond Blume-Capel model
at equilibrium have been constructed in
Refs.~\cite{Malakis1,Malakis2} and, more recently, a dedicated
numerical study at the first-order transition regime revealed that
the induced under disorder continuous transition belongs to the
universality class of the random Ising model with logarithmic
corrections~\cite{Fytas17}.

\section{Observables}
\label{sec:observables}

In order to determine the universality aspects of the kinetic
random-bond Ising and Blume-Capel models, we shall consider the
half-period dependencies of various thermodynamic observables. The
main quantity of interest is the period-averaged magnetization
\begin{equation}\label{eq:5}
Q=\frac{1}{2t_{1/2}}\oint M(t)dt,
\end{equation}
where the integration is performed over one cycle of the
oscillating field. Given that for finite systems in the
dynamically ordered phase the probability density of $Q$ becomes
bimodal, one has to measure the average norm of $Q$ in order to
capture symmetry breaking, so that $\langle |Q| \rangle$ defines
the dynamic order parameter of the system. In the above
Eq.~(\ref{eq:5}), $M(t)$ is the time-dependent magnetization per
site
\begin{equation}\label{eq:6}
 M(t)=\frac{1}{N}\sum_{x = 1}^{N}\sigma_{x}(t),
\end{equation}
where $N = L\times L$ defines the total number of spins and $L$ the linear dimension of the lattice.

To characterize and quantify the transition using finite-size
scaling arguments we must also define quantities analogous to the
susceptibility in equilibrium systems. The scaled variance of the
dynamic order parameter
\begin{equation}\label{eq:7}
\chi_{L}^{Q} = N\left[\langle Q^2\rangle_{L} -\langle |Q|
\rangle^2_{L} \right],
\end{equation}
has been suggested as a proxy for the nonequilibrium
susceptibility, also theoretically justified via
fluctuation-dissipation relations~\cite{Robb1}. Similarly, one may
also measure the scaled variance of the period-averaged energy
\begin{equation} \label{eq:8}
\chi_{L}^{E} = N\left[\langle E^2\rangle_{L} -\langle E
\rangle^2_{L} \right],
\end{equation}
so that $\chi_{L}^{E}$ can be considered as the corresponding heat capacity. 
Here $E$ denotes the
cycle-averaged energy corresponding to the cooperative part of the
Hamiltonians~(\ref{eq:1}) and (\ref{eq:2}). With the help of the
dynamic order parameter $Q$ we may define the fourth-order Binder
cumulant~\cite{Sides1,Sides2}
\begin{equation}
\label{eq:9} U_{L}=1-\frac{\langle |Q|^4\rangle_L}{3\langle |Q|^2
\rangle_L^2},
\end{equation}
a very useful observable for the characterization of universality
classes~\cite{Binder81}.

\section{Simulation Details}
\label{sec:numerics}

We performed Monte Carlo simulations on square lattices with
periodic boundary conditions using the single-site update
Metropolis algorithm~\cite{Metropolis,Binder,Newman}. This
approach, together with the stochastic Glauber
dynamics~\cite{Glauber:63}, consists the standard recipe in
kinetic Monte Carlo simulations~\cite{Buendia3}. Let us briefly outline below the steps of our computer algorithm:
\begin{enumerate}

\item A lattice site is selected randomly among the $L\times L$ options.

\item The spin  variable located at the selected site 
is flipped, keeping the other spins in the system fixed. 

\item The energy change originating from this spin flip operation is calculated using the Hamiltonians of Eqs.~(\ref{eq:1}) of (\ref{eq:2}) as follows: $\Delta \mathcal{H}=\mathcal{H}_{\rm a}-\mathcal{H}_{\rm o}$, 
where $\mathcal{H}_{\rm a}$ denotes the energy of 
the system after the trial switch of the selected spin and $\mathcal{H}_{\rm o}$ corresponds to the total 
energy of the system with the old spin configuration. The probability to accept the proposed spin update is given by:
\begin{equation}
\label{eq:10}
 W_{M}\left(\sigma_{x}\rightarrow \sigma_{x}' \right)=
\begin{cases}
   \exp(-\Delta \mathcal{H}/k_{\rm B}T)        & \text{if } \mathcal{H}_{a} \geq \mathcal{H}_{o} \\
   1        & \text{if } \mathcal{H}_{\rm a} < \mathcal{H}_{\rm o}.
  \end{cases}
\end{equation}

\item If the energy is lowered, the spin flip is always accepted. 

\item If the energy is increased, a random number $R$ is generated, such that $0 < R < 1$: If this number $R$ is less than or equal to the calculated Metropolis transition probability the selected spin is flipped. Otherwise, the old spin configuration remains unchanged. 

\end{enumerate}

\begin{figure}[t]
\centering
\includegraphics[width=8 cm]{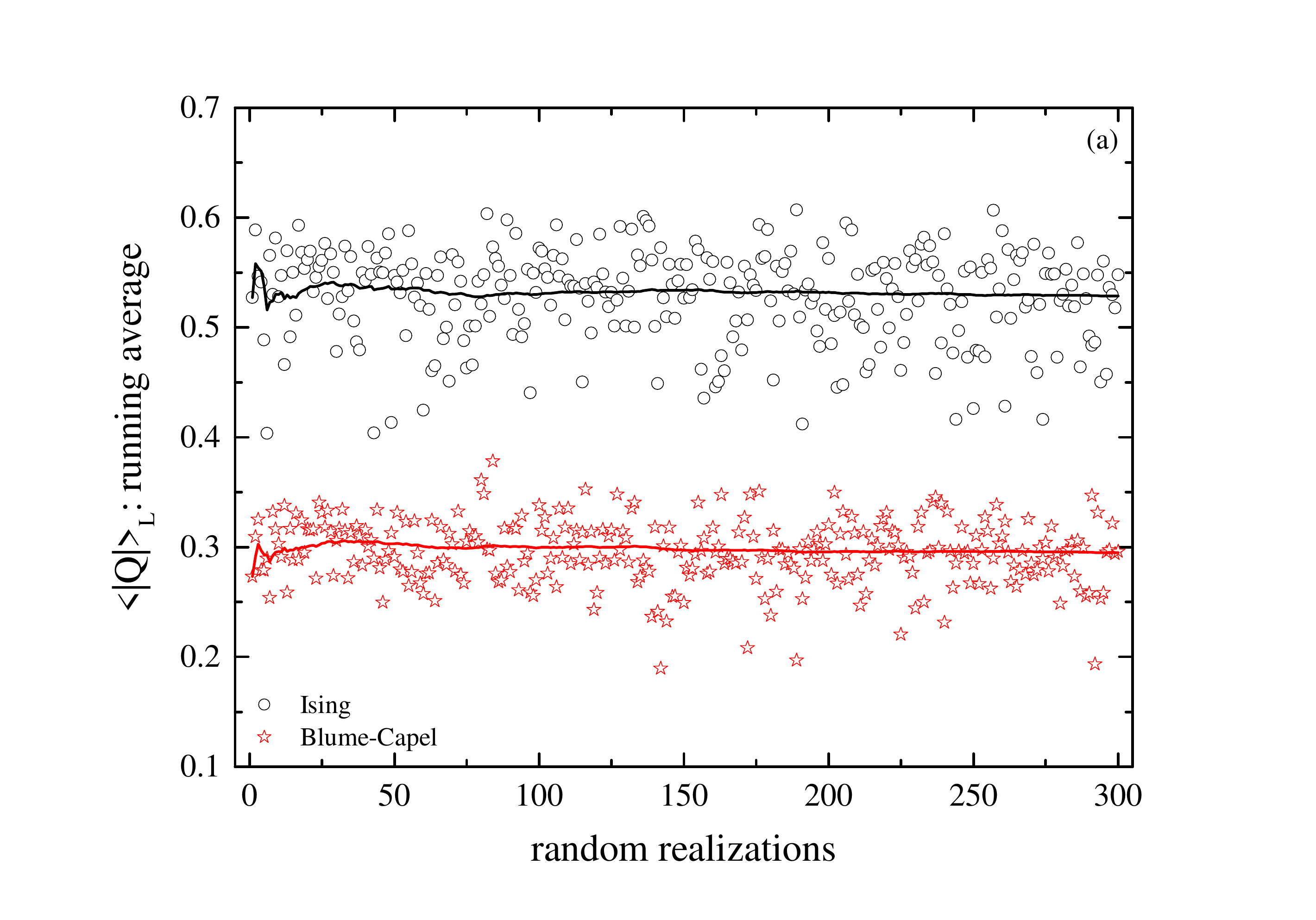}\\
\includegraphics[width=8 cm]{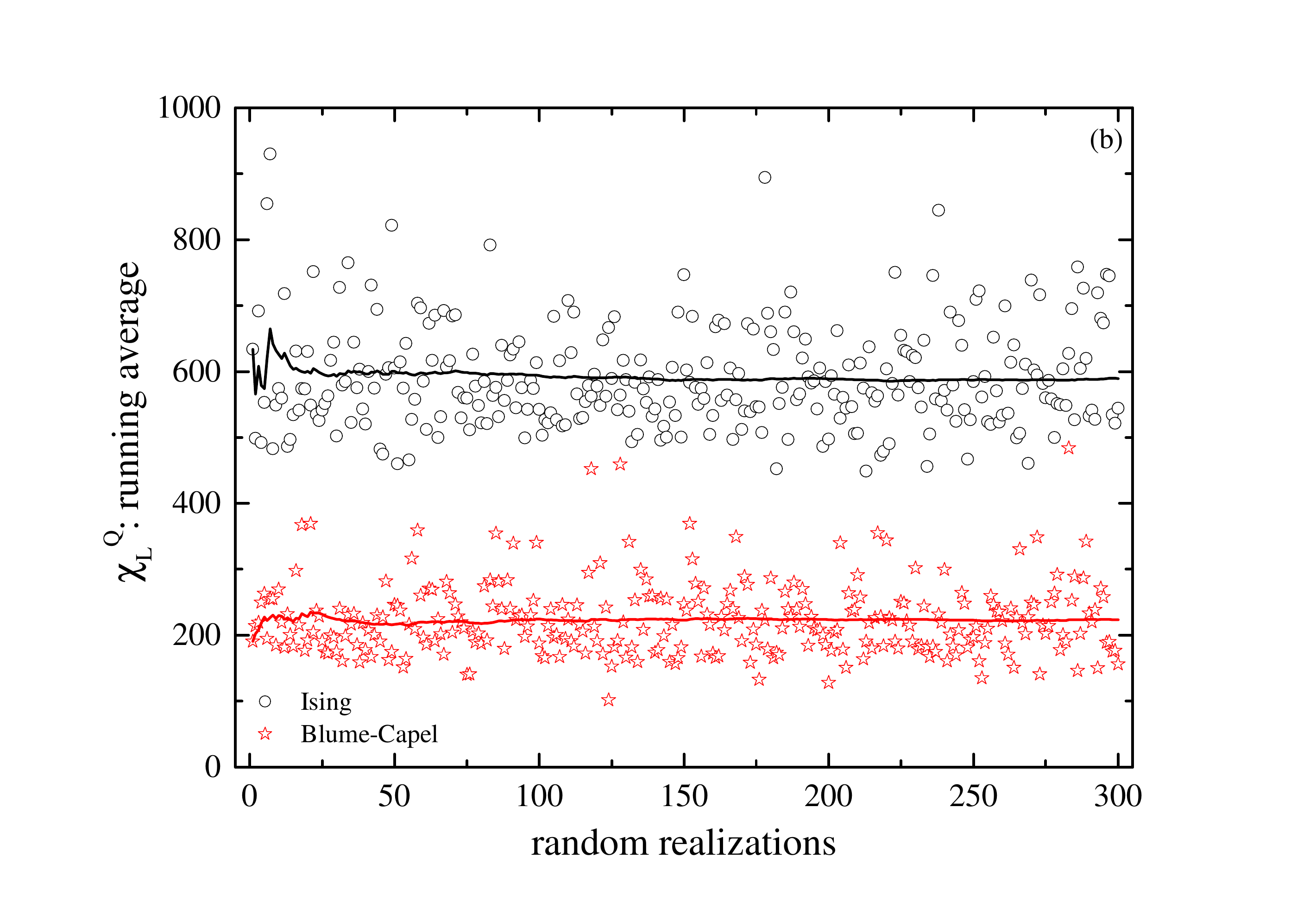}\\
\includegraphics[width=8 cm]{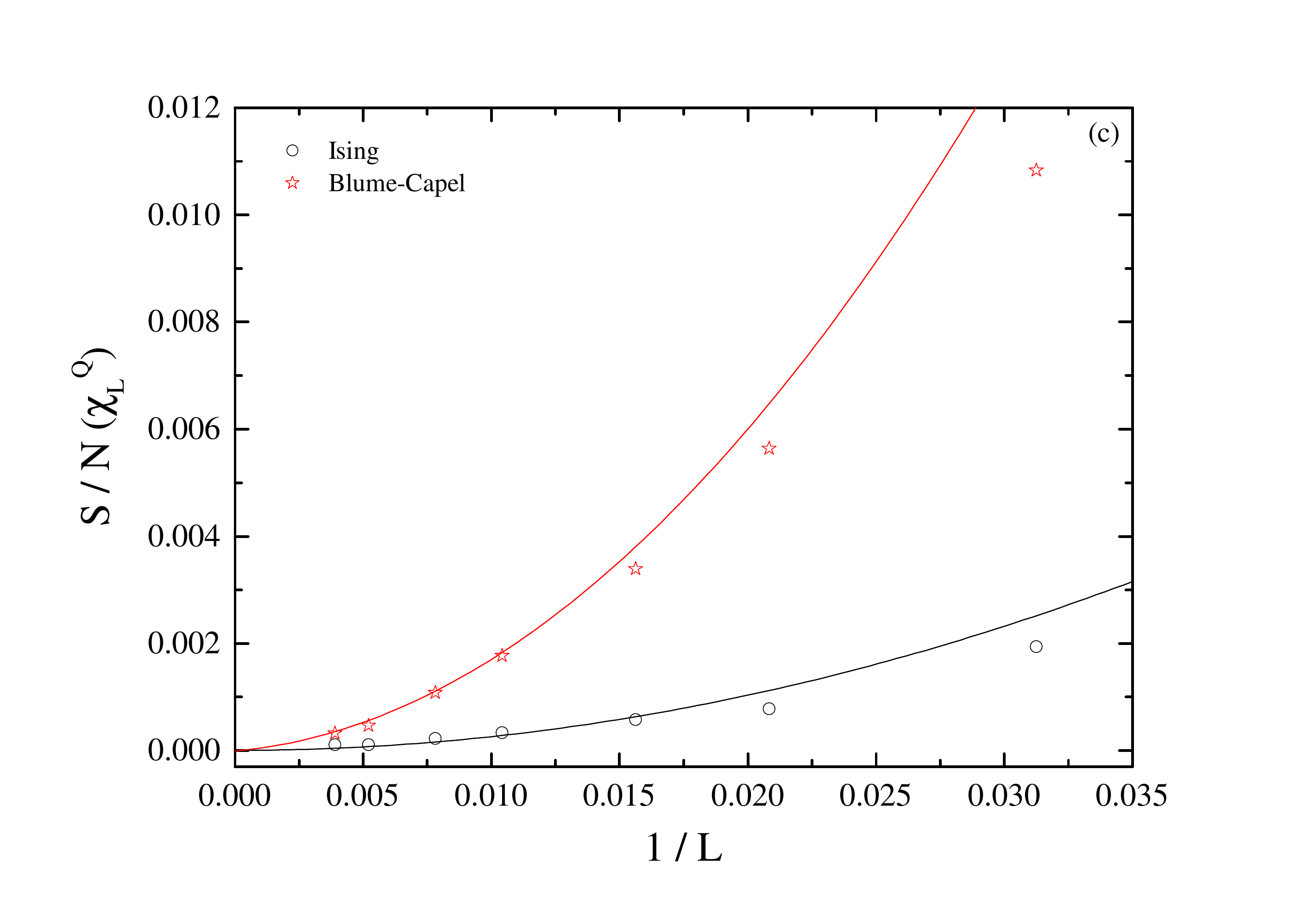}
\caption{\label{fig:run_average} Disorder distributions of the nonequilibrium dynamic order parameter, panel(a), and susceptibility maxima, panel (b), for a lattice size $L = 128$ and for both models considered in this work. The running averages over the samples are shown by the solid lines. Panel (c) shows the signal-to-noise ratio $S/N$ of the dynamic susceptibility, that is the ratio of the relative variance of the distribution over the square of its mean value, as a function of the inverse linear size. The solid lines are second-order polynomial fittings to $1/L$ for the larger system sizes. Clearly, $S/N(\chi_{L}^{Q})\rightarrow 0$ as $L\rightarrow \infty$, indicating that self-averaging is restored in the thermodynamic limit for both kinetic disordered systems~\cite{Aharony96,Wiseman98}.}
\end{figure}

Using the above scheme we simulated system sizes within the range $L  = 32 - 256$. For each system size $300$
independent realizations of the disorder have been generated -- see
Fig.~\ref{fig:run_average} for characteristic illustrations of disorder averages and their relative variance -- and for each random sample the
following simulation protocol has been used: The first $10^{3}$
periods of the external field have been discarded during the
thermalization process and numerical data were collected and
analyzed during the following $10^{4}$ periods of the field. The
time unit in our simulations was one Monte Carlo step per site
(MCSS) and error bars have been estimated using the jackknife
method~\cite{Newman}. To give a flavor of the actual CPU time of our computations we note that the simulation times needed for a single disorder realization of the kinetic Ising model on a single node of a Dual Intel Xeon E5-2690 V4 processor were 6 hours for $L = 32$ and 11 days for $L = 256$. The analogous CPU times for the kinetic Blume-Capel model were 3 hours and 9 days for $L = 32$ and $L = 256$, respectively. For the Ising model we fixed the value of the disorder strength to $r = 1/7$, whereas for the Blume-Capel model we focused on the value $\Delta = 1.975$ in the originally first-order regime selecting now $r = 0.75/1.25$ following Refs.~\cite{Malakis2,Malakis3}. Appropriate choices of the magnetic-field
strength, $h_{0} = 0.3$, and the temperature, $T^{\rm (Ising)} =
0.8\times T_{\rm c}^{\rm (Ising)}$ and $T^{\rm (BC)} = 0.6\times
T_{\rm c}^{\rm (BC)}$, ensured that the system lies in the
multi-droplet regime~\cite{Park}. Here, $T_{\rm c}^{\rm (Ising)} =
1.7781$~\cite{Fisch,Kinzel} and $T_{\rm c}^{\rm (BC)} =
0.626$~\cite{Malakis2,Malakis3} are the equilibrium critical
temperatures of the Ising and Blume-Capel models for the
particular choices of $r$ and $\Delta$ considered in this work.

For the fitting process on the numerical data
we restricted ourselves to data with $L \geq L_{\rm min}$. As
usual, to determine an acceptable $L_{\rm min}$ we employed the
standard $\chi^{2}$-test of goodness of fit~\cite{Press}.
Specifically, the $p$-value of our $\chi^{2}$-test is the
probability of finding an $\chi^{2}$ value which is even larger
than the one actually found from our data. We considered a fit as
being fair only if $10\% < p < 90\%$.

\section{Results and discussion}
\label{sec:results}

\begin{figure}[t]
\centering
\includegraphics[width=8 cm]{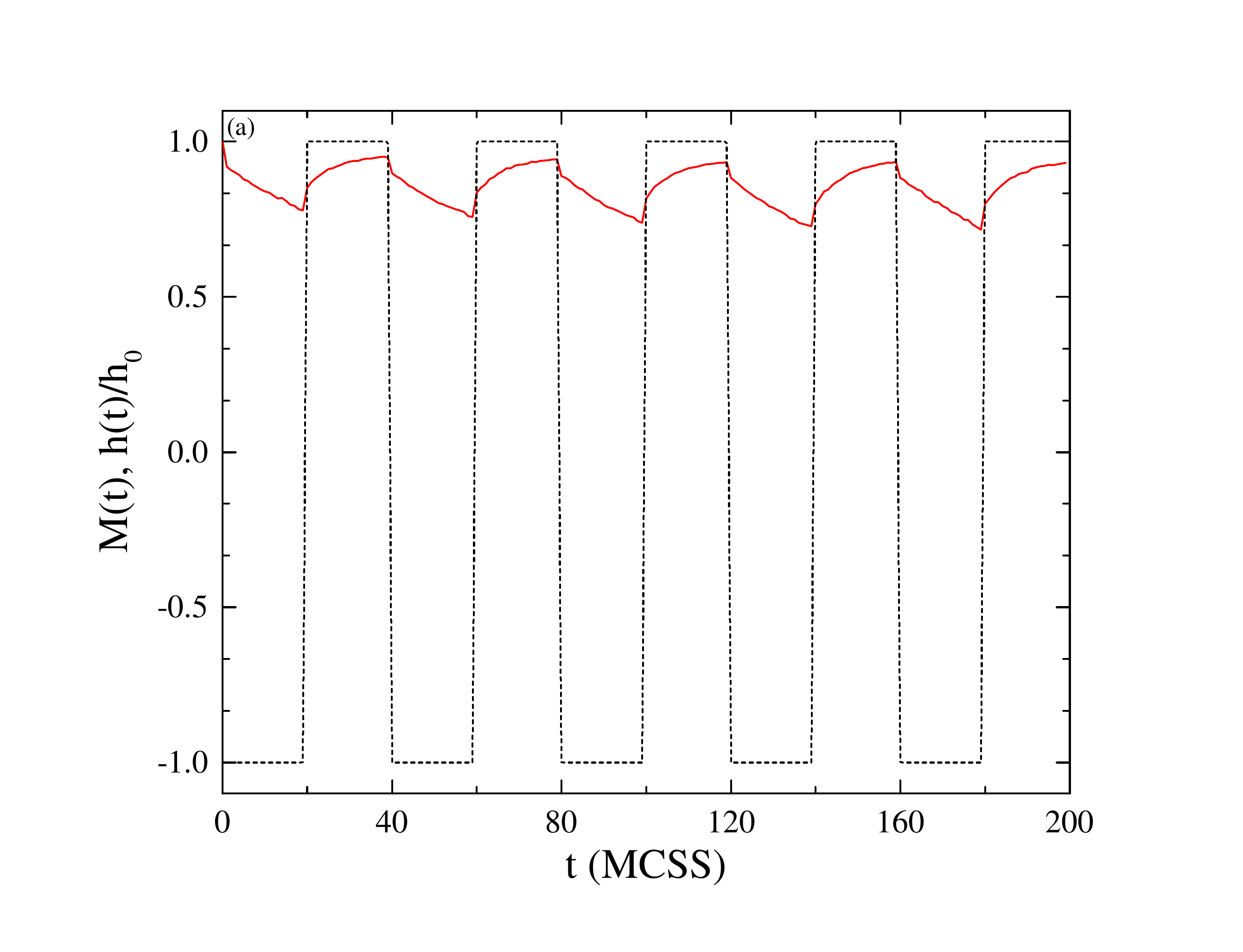}\\
\includegraphics[width=8 cm]{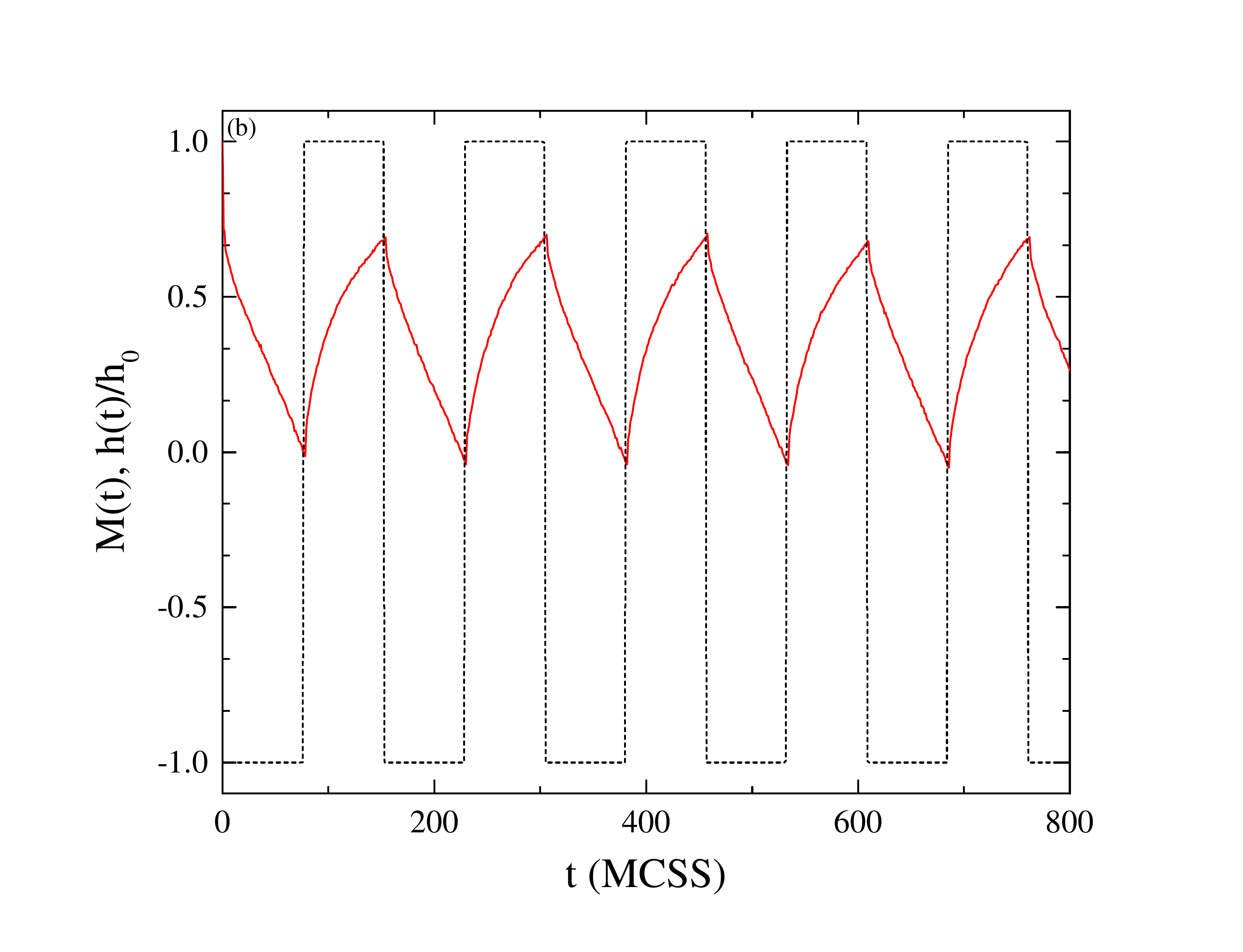}\\
\includegraphics[width=8 cm]{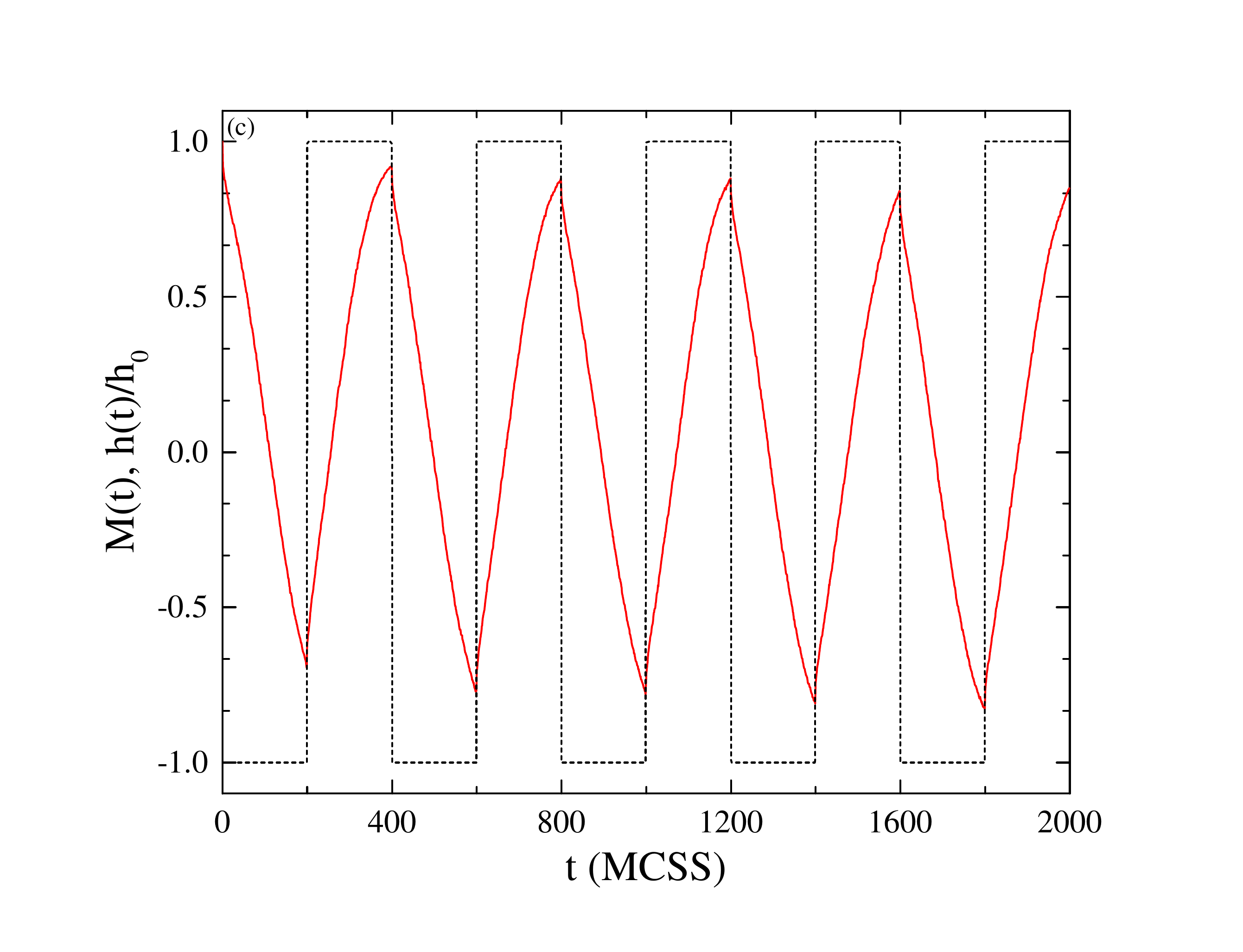}
\caption{\label{fig:time_series_Ising} Time series of the
magnetization (red solid curves) of the kinetic random-bond Ising
model under the presence of a square-wave magnetic field (black
dashed lines) for $L = 96$ and three values of the half period of
the external field: (a) $t_{1/2} = 20$ MCSS, corresponding to a
dynamically ordered phase, (b) $t_{1/2} = 76$ MCSS, close to the
dynamic phase transition, and (c) $t_{1/2} = 200$ MCSS,
corresponding to a dynamically diordered phase. Note that for the
sake of clarity the ratio $h(t)/h_{0}$ is displayed.}
\end{figure}

\begin{figure}[t]
\includegraphics[width=10 cm]{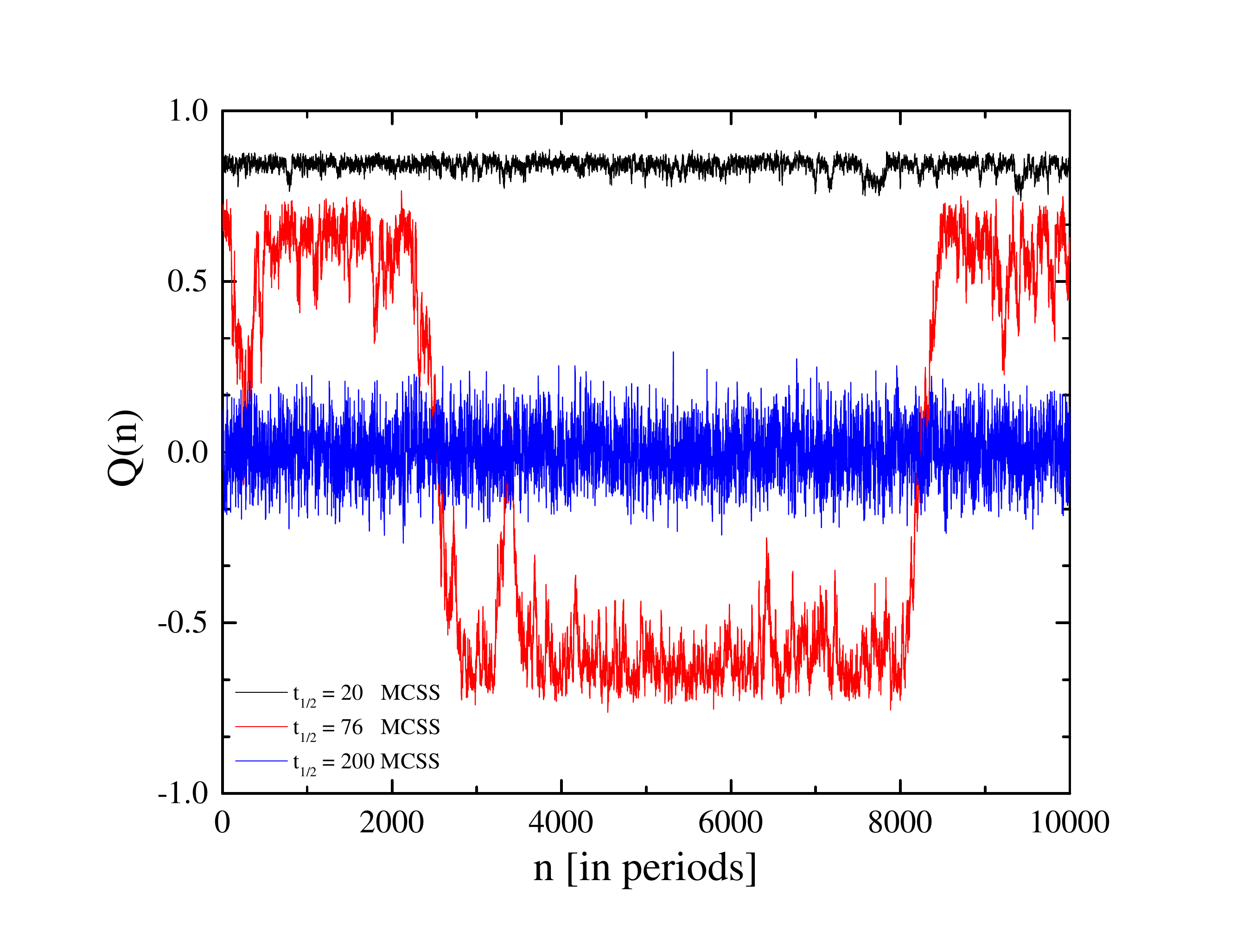}
\caption{\label{fig:series_Ising} Period dependencies of the
dynamic order parameter of the kinetic random-bond Ising model for
$L = 96$. Results are shown for the three characteristic cases of
the half period of the external field, following
Fig.~\ref{fig:time_series_Ising}.}
\end{figure}

\begin{figure}[t]
\centering
\includegraphics[width=8 cm]{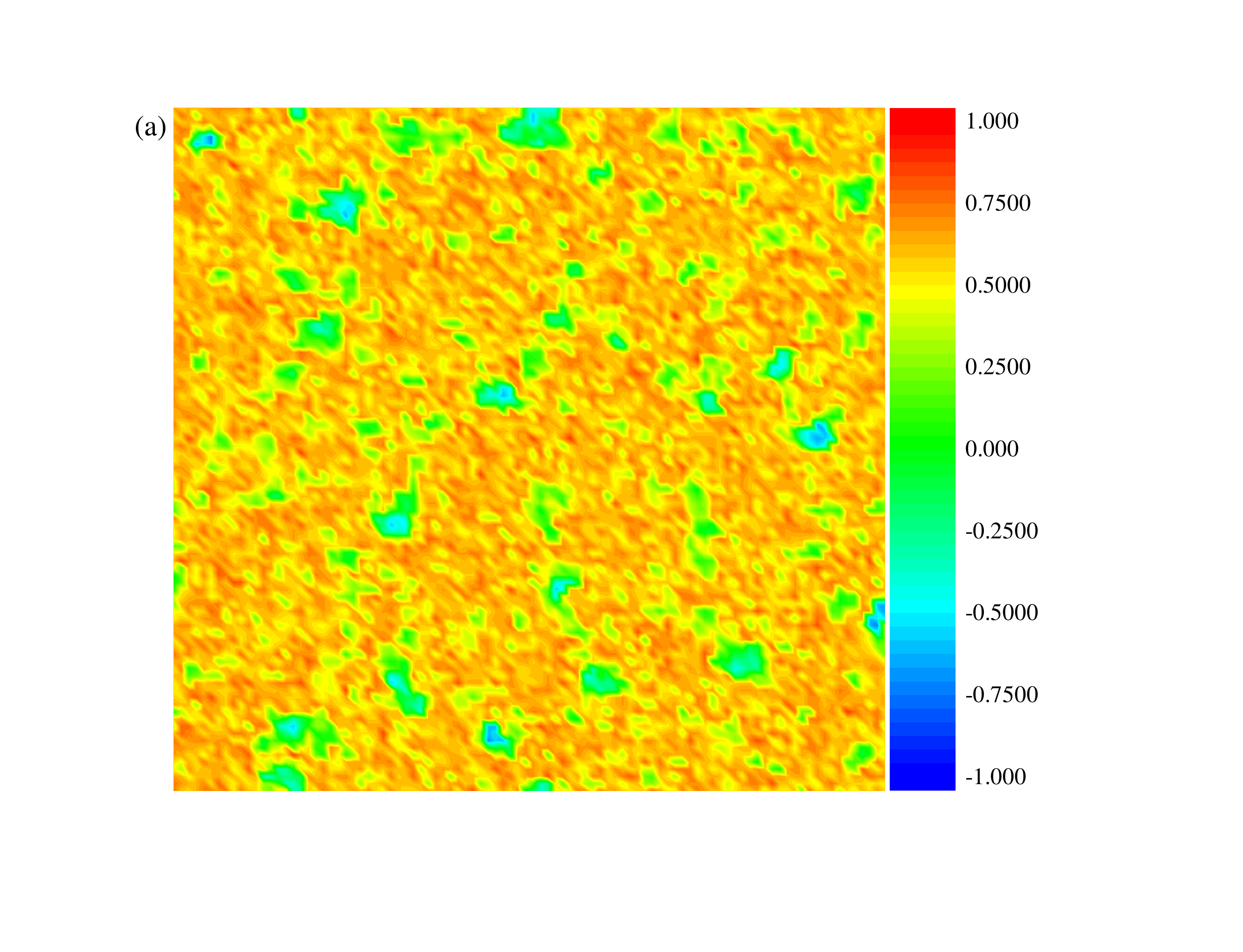}\\
\includegraphics[width=8 cm]{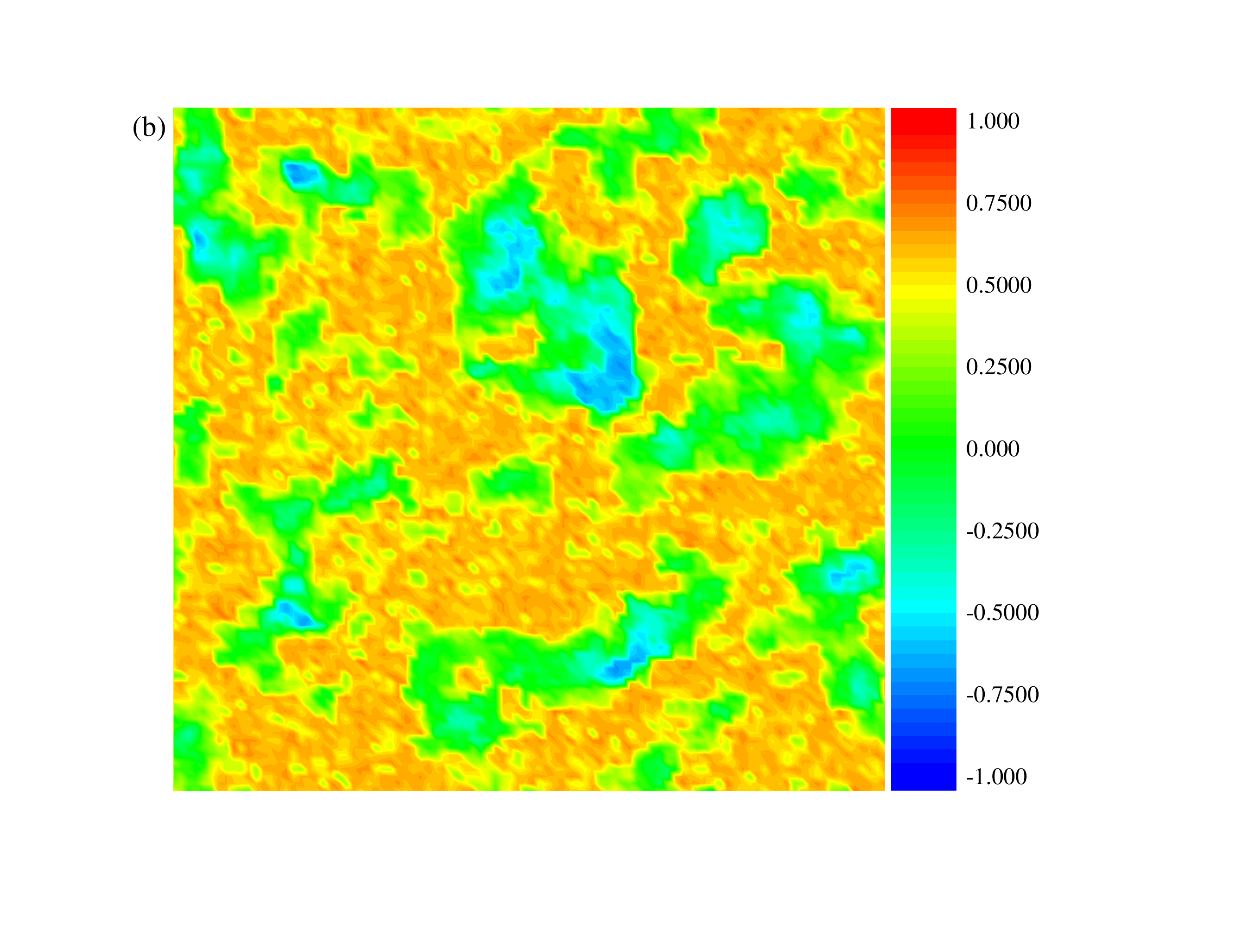}\\
\includegraphics[width=8 cm]{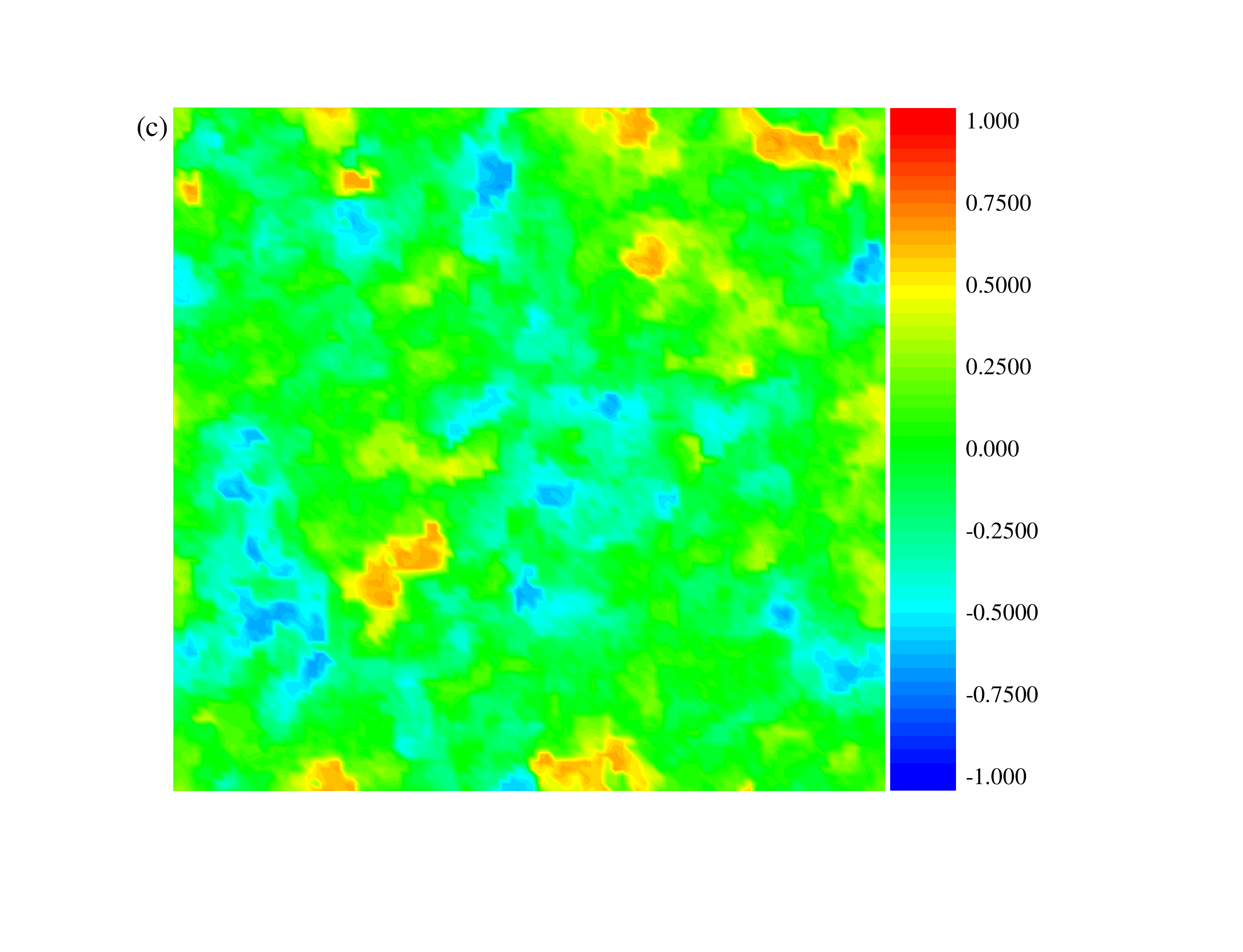}
\caption{\label{fig:local_Ising} Configurations of the local
dynamic order parameter $\{Q_{x}\}$ of the random-bond kinetic
Ising model for $L = 96$. The ``snapshots'' of $\{Q_{x}\}$ for
each regime are the set of local period-averaged spins during some
representative period. Three panels are shown: (a) $t_{1/2} = 20$
MCSS $ < t_{1/2}^{\rm c}$ -- dynamically ordered phase, (b)
$t_{1/2} = 76$ MCSS $\approx t_{1/2}^{\rm c}$ -- near the dynamic
phase transition, and (c) $t_{1/2} = 200$ MCSS $ > t_{1/2}^{\rm
c}$ -- dynamically disordered phase.}
\end{figure}

\begin{figure}[t]
\centering
\includegraphics[width=8 cm]{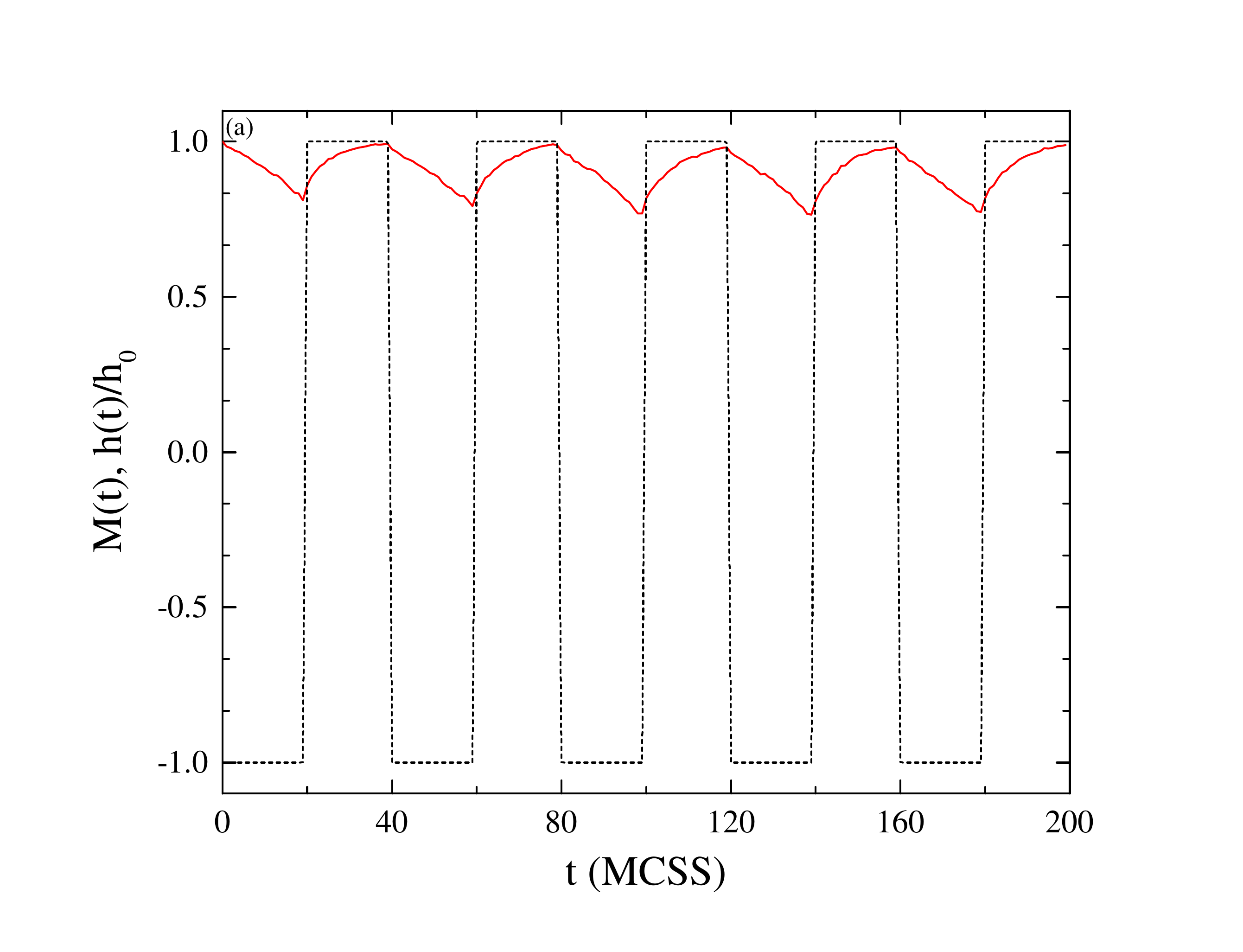}\\
\includegraphics[width=8 cm]{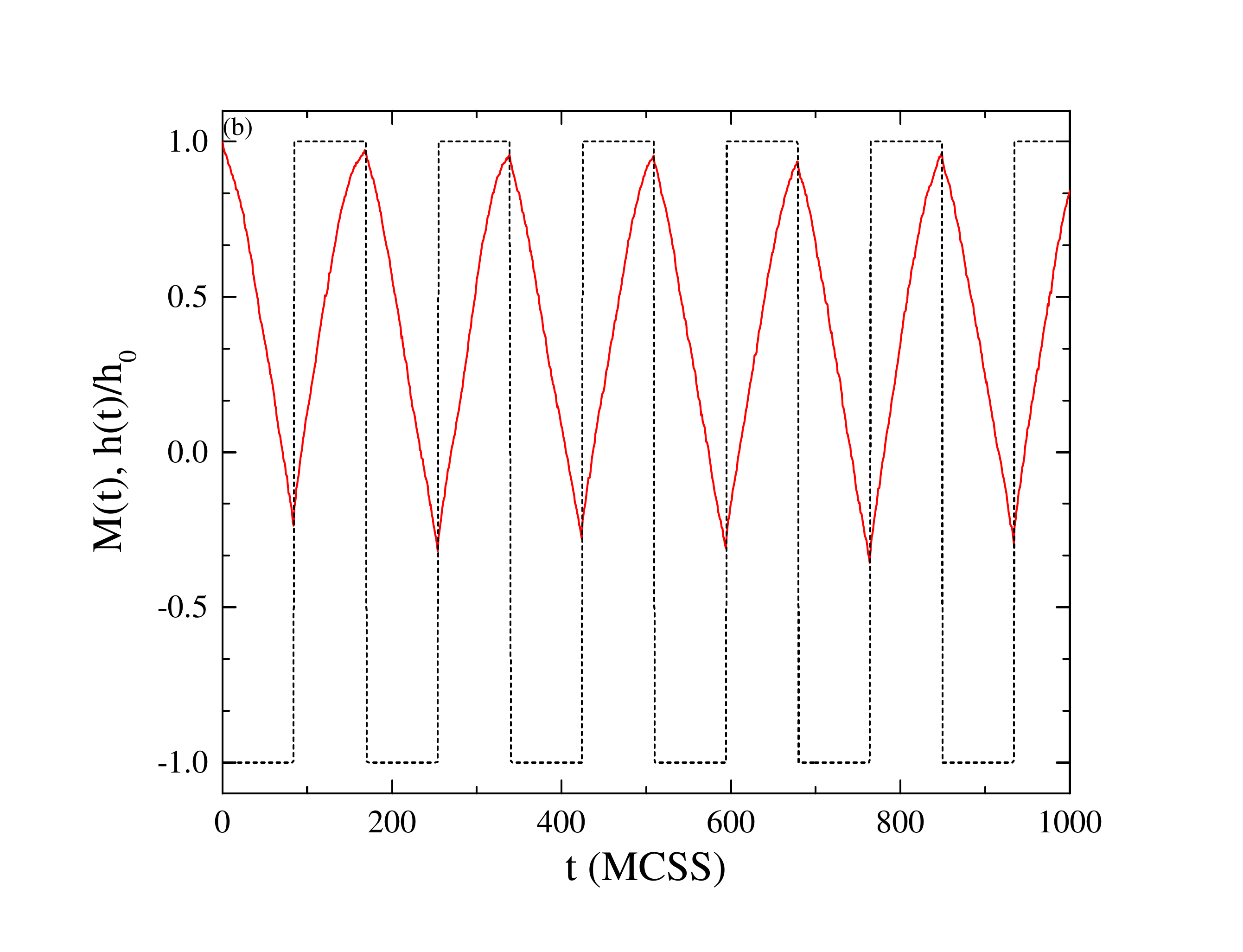}\\
\includegraphics[width=8 cm]{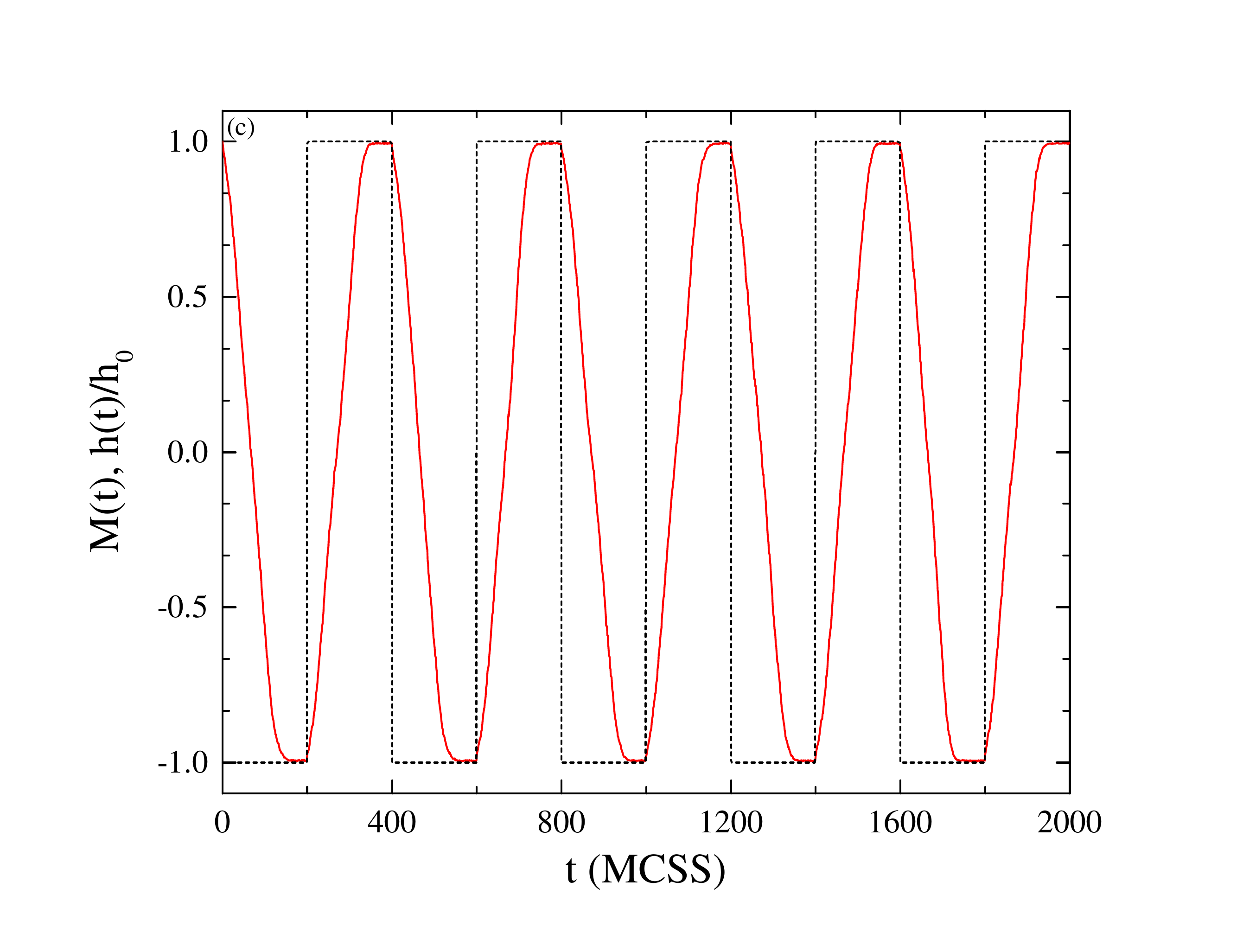}
\caption{\label{fig:time_series_BC} Time series of the
magnetization (red solid curves) of the kinetic random-bond
$\Delta = 1.975$ Blume-Capel model under the presence of a
square-wave magnetic field (black dashed lines) for $L = 96$ and
three values of the half period of the external field: (a)
$t_{1/2} = 20$ MCSS, corresponding to a dynamically ordered phase,
(b) $t_{1/2} = 85$ MCSS, close to the dynamic phase transition,
and (c) $t_{1/2} = 200$ MCSS, corresponding to a dynamically
diordered phase. Note that for the sake of clarity the ratio
$h(t)/h_{0}$ is displayed.}
\end{figure}

\begin{figure}[t]
\includegraphics[width=10 cm]{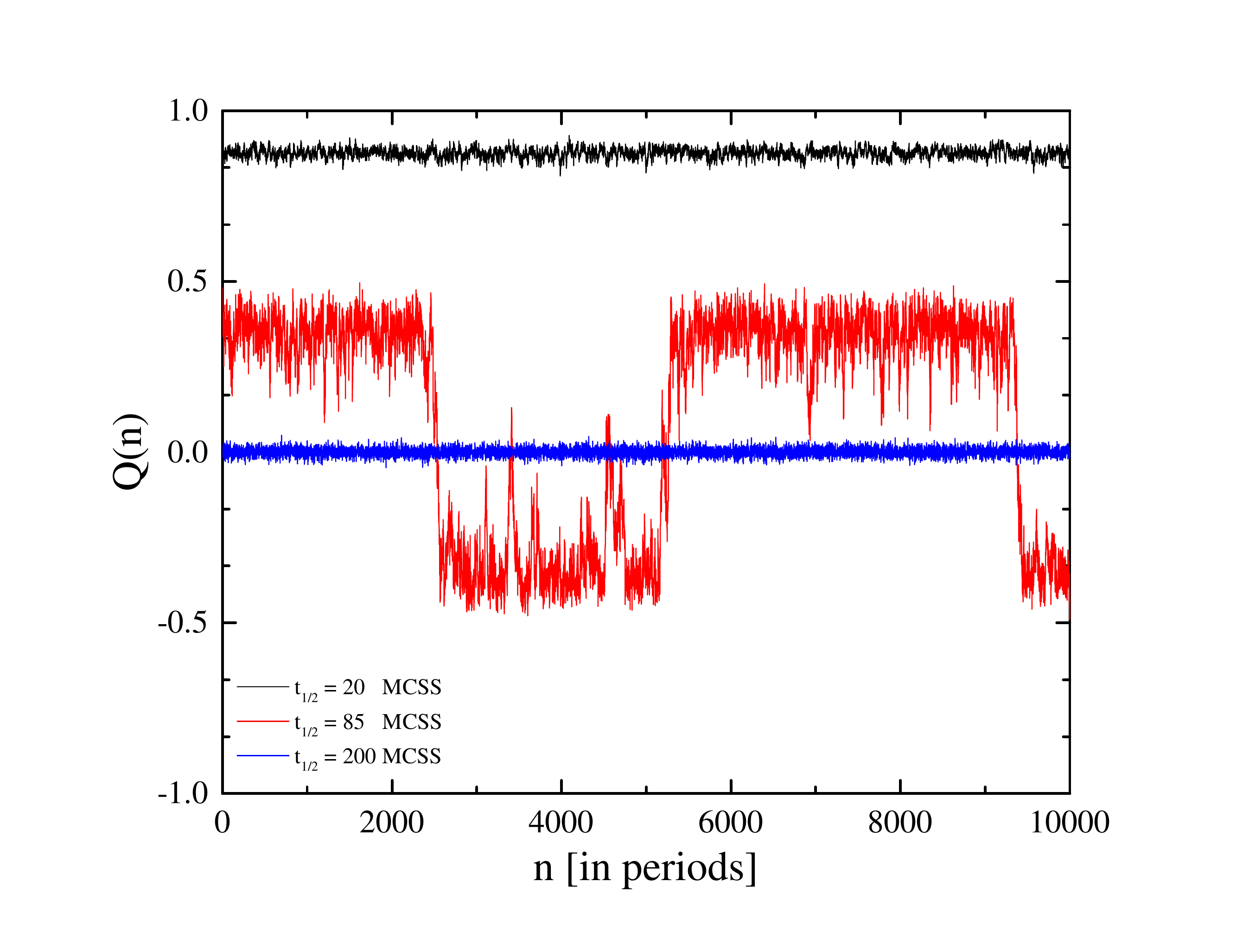}
\caption{\label{fig:series_BC} Period dependencies of the dynamic
order parameter of the kinetic random-bond $\Delta = 1.975$
Blume-Capel model for $L = 96$. Results are shown for the three
characteristic cases of the half period of the external field,
following Fig.~\ref{fig:time_series_BC}.}
\end{figure}

\begin{figure}[t]
\centering
\includegraphics[width=8 cm]{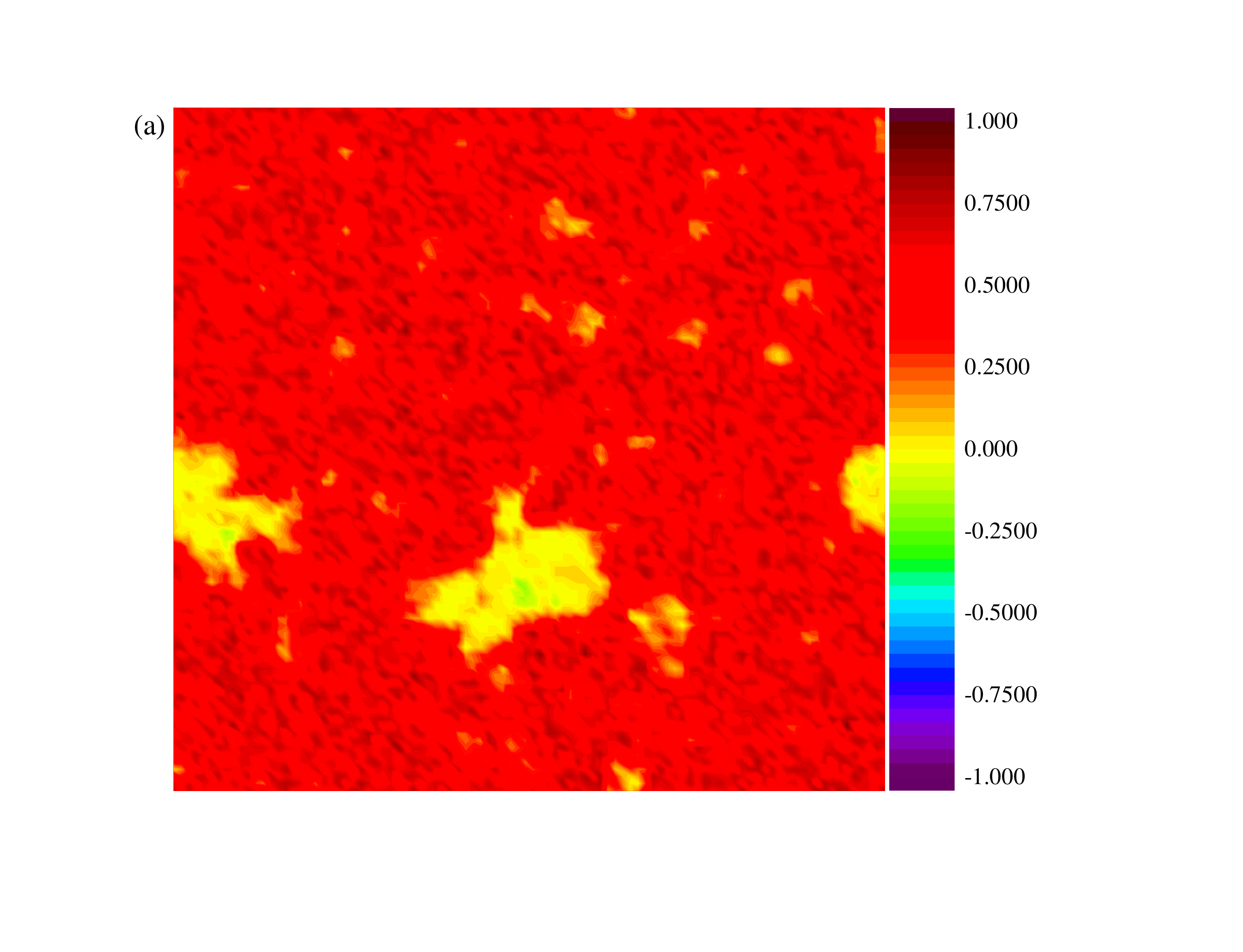}\\
\includegraphics[width=8 cm]{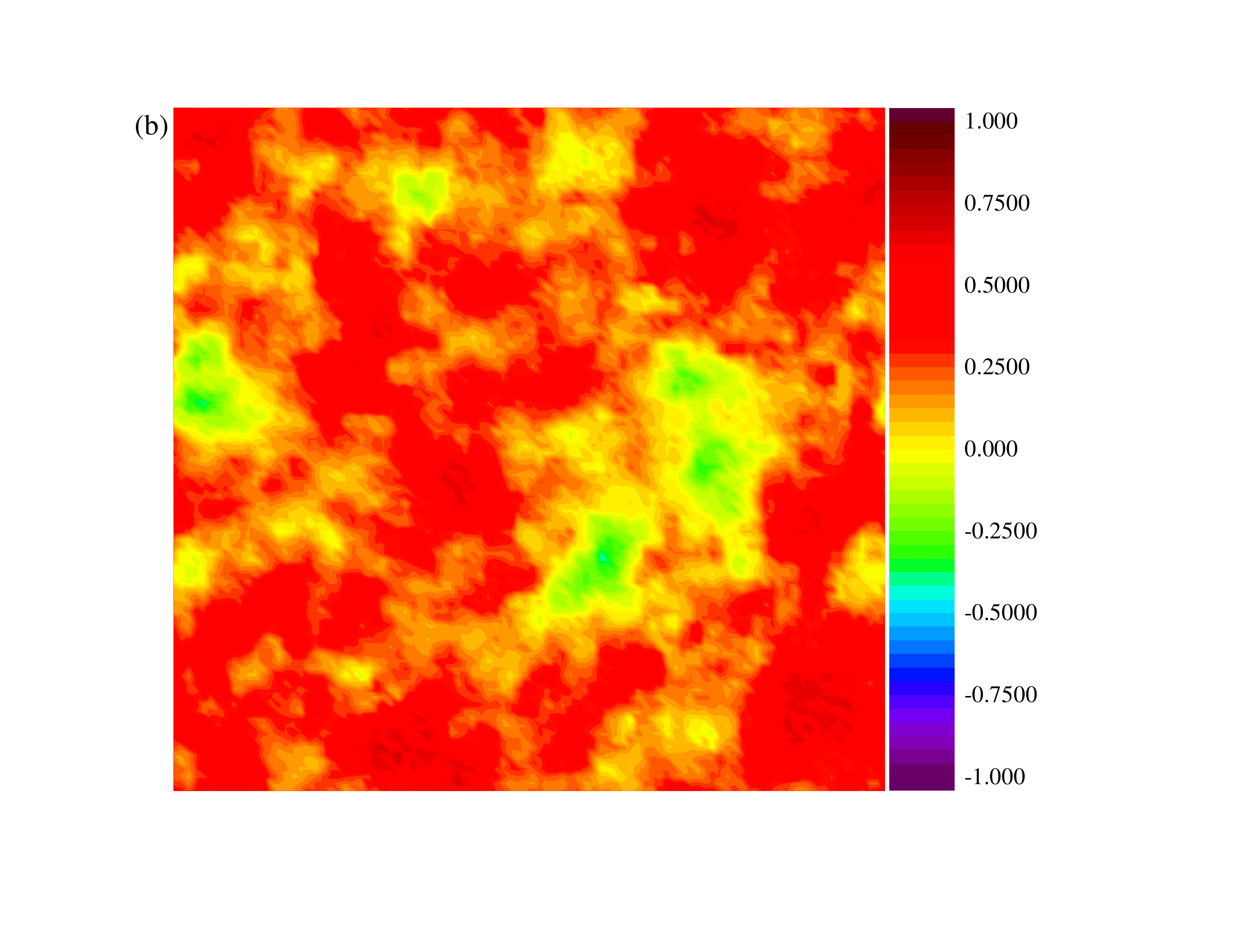}\\
\includegraphics[width=8 cm]{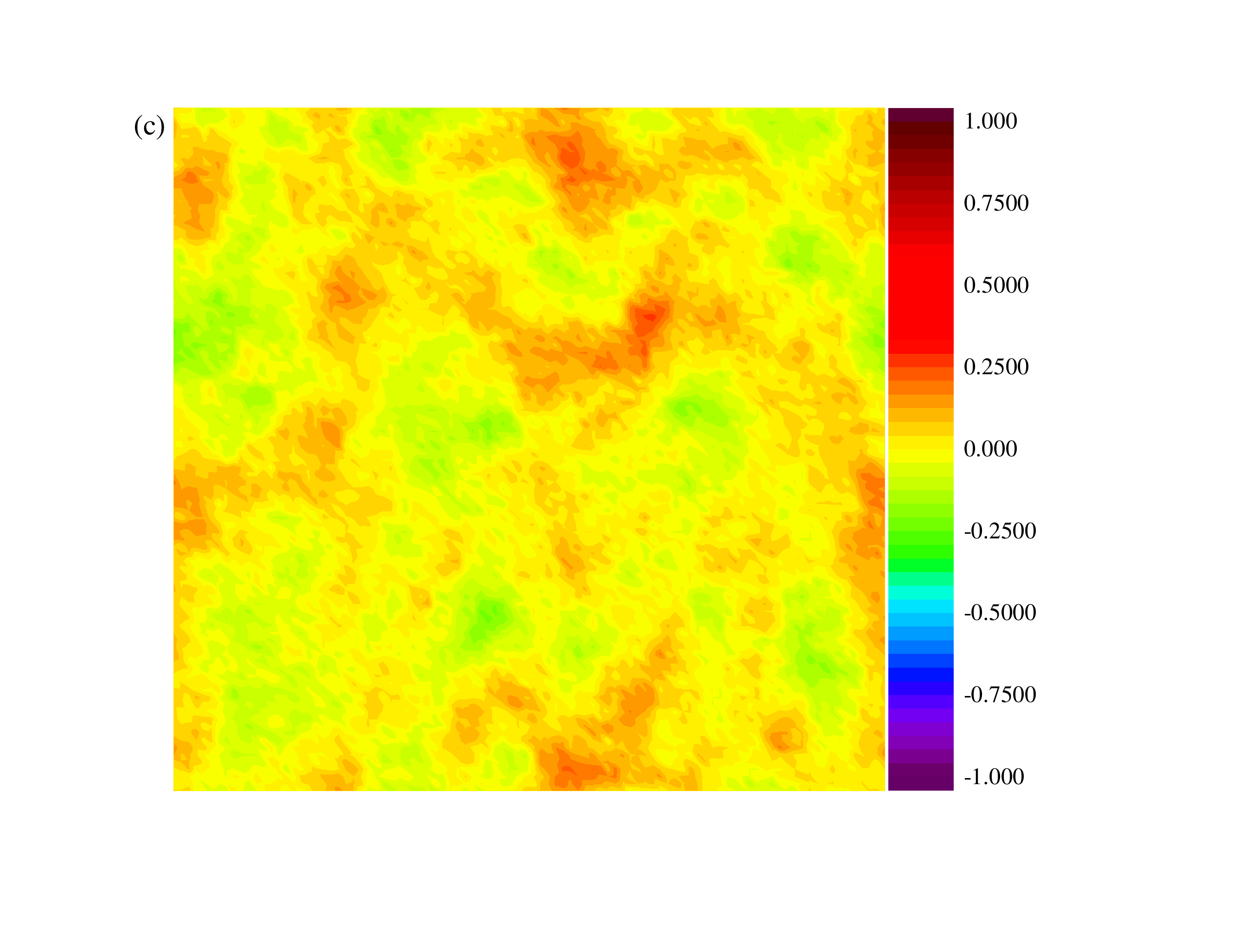}
\caption{\label{fig:local_BC} Configurations of the local dynamic
order parameter $\{Q_{x}\}$ of the random-bond kinetic $\Delta =
1.975$ Blume-Capel for $L = 96$. The ``snapshots'' of $\{Q_{x}\}$
for each regime are the set of local period-averaged spins during
some representative period. Three panels are shown: (a) $t_{1/2} =
20$ MCSS $ < t_{1/2}^{\rm c}$ -- dynamically ordered phase, (b)
$t_{1/2} = 85$ MCSS $\approx t_{1/2}^{\rm c}$ -- near the dynamic
phase transition, and (c) $t_{1/2} = 200$ MCSS $ > t_{1/2}^{\rm
c}$ -- dynamically disordered phase.}
\end{figure}

\begin{figure}[t]
\centering
\includegraphics[width=8 cm]{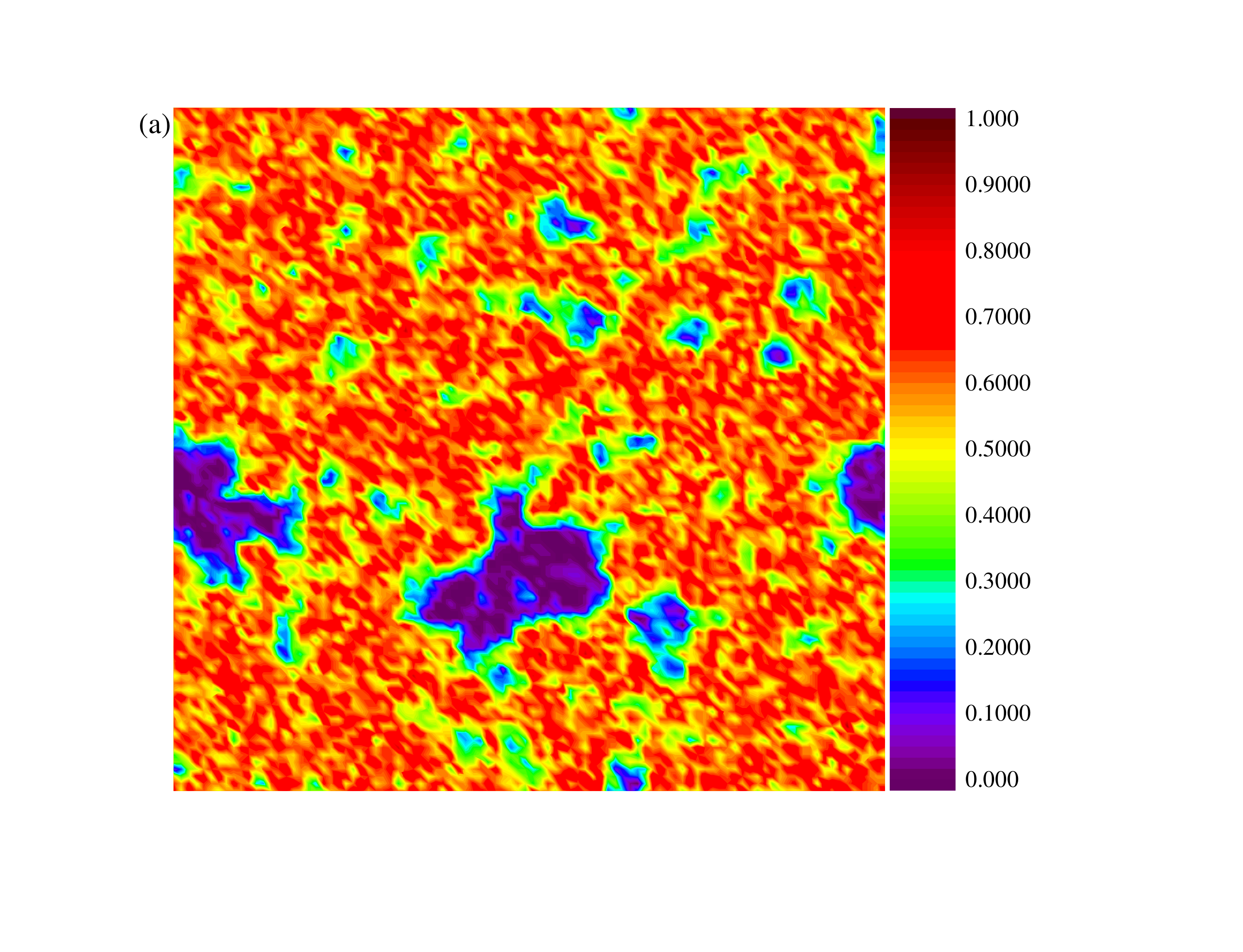}\\
\includegraphics[width=8 cm]{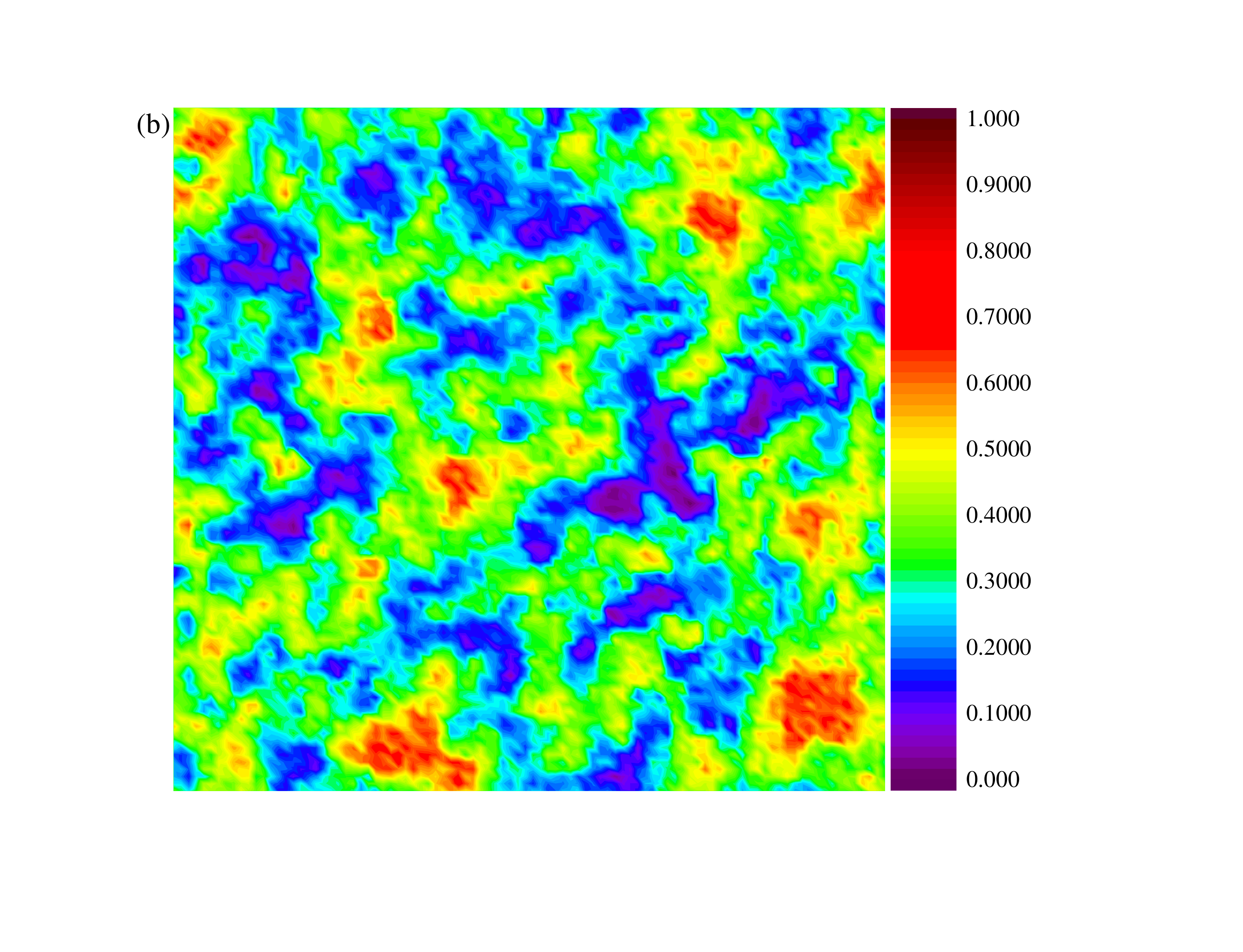}\\
\includegraphics[width=8 cm]{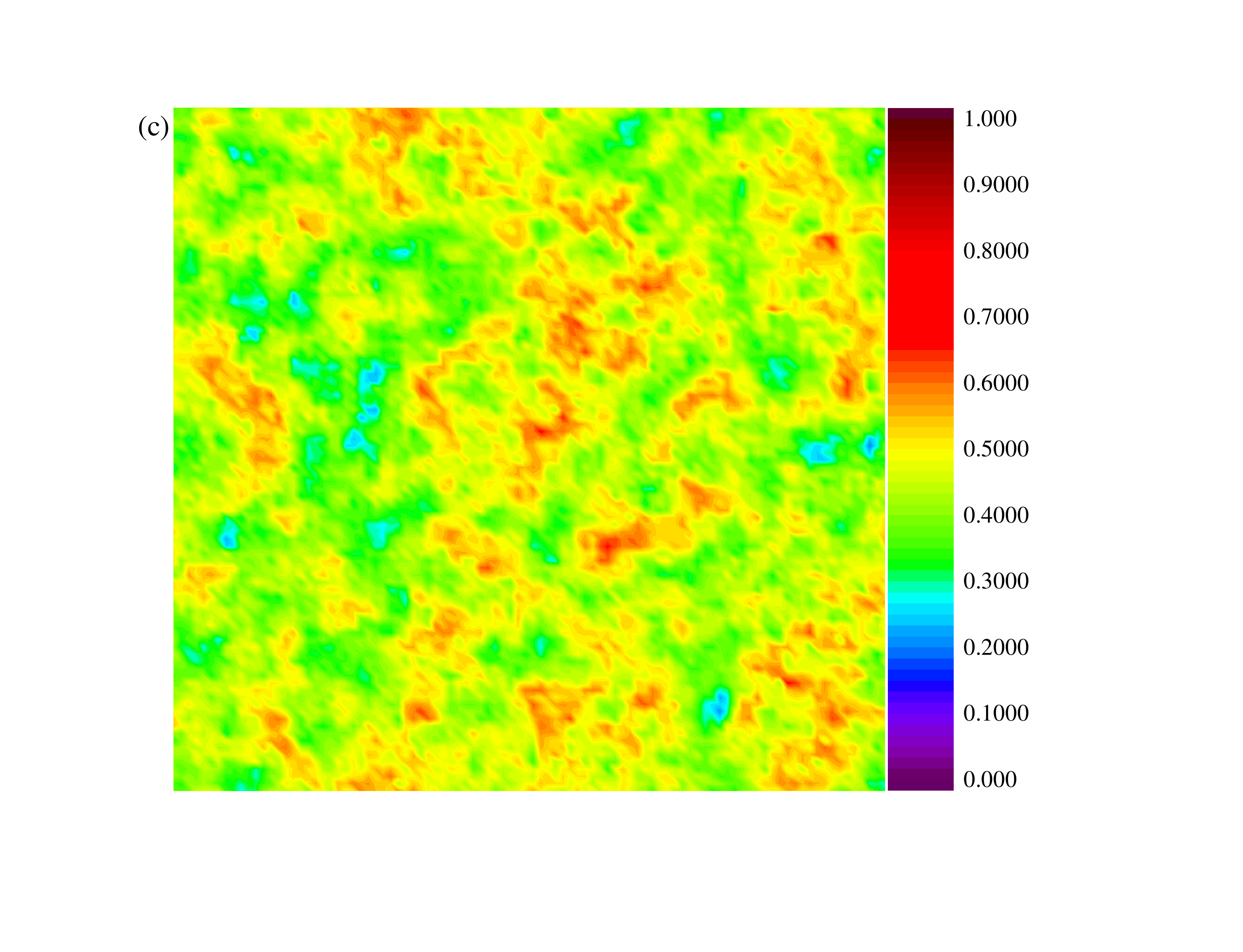}
\caption{\label{fig:quad_BC} In full analogy to
Fig.~\ref{fig:local_BC} we show snapshots of the period-averaged
dynamic quadrupolar moment conjugate to the crystal-field coupling
$\Delta$. The simulation parameters are exactly the same to those
used in Fig.~\ref{fig:local_BC} for all three panels (a) - (c).}
\end{figure}

As a starting point let us describe shortly the mechanism
underlying the dynamical ordering in kinetic ferromagnets (here,
under the presence of quenched randomness), as exemplified in
Figs.~\ref{fig:time_series_Ising} - \ref{fig:local_Ising} for the
Ising model, and Figs.~\ref{fig:time_series_BC} -
\ref{fig:quad_BC} for the Blume-Capel model. In both cases,
results for a single realization of the disorder are shown over a
system size of $L = 96$.

\begin{figure}[t]
\centering
\includegraphics[width=10 cm]{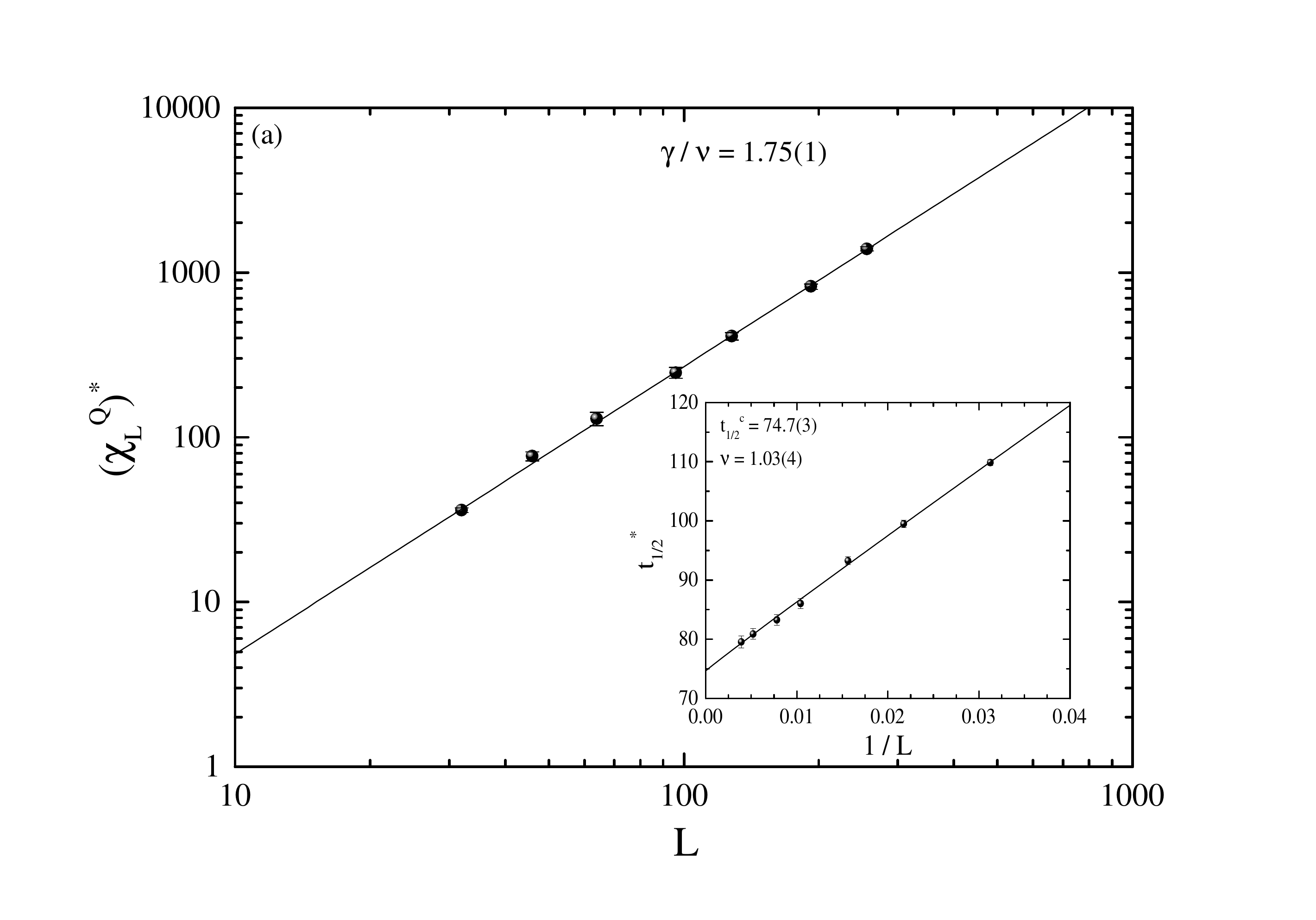}\\
\includegraphics[width=10 cm]{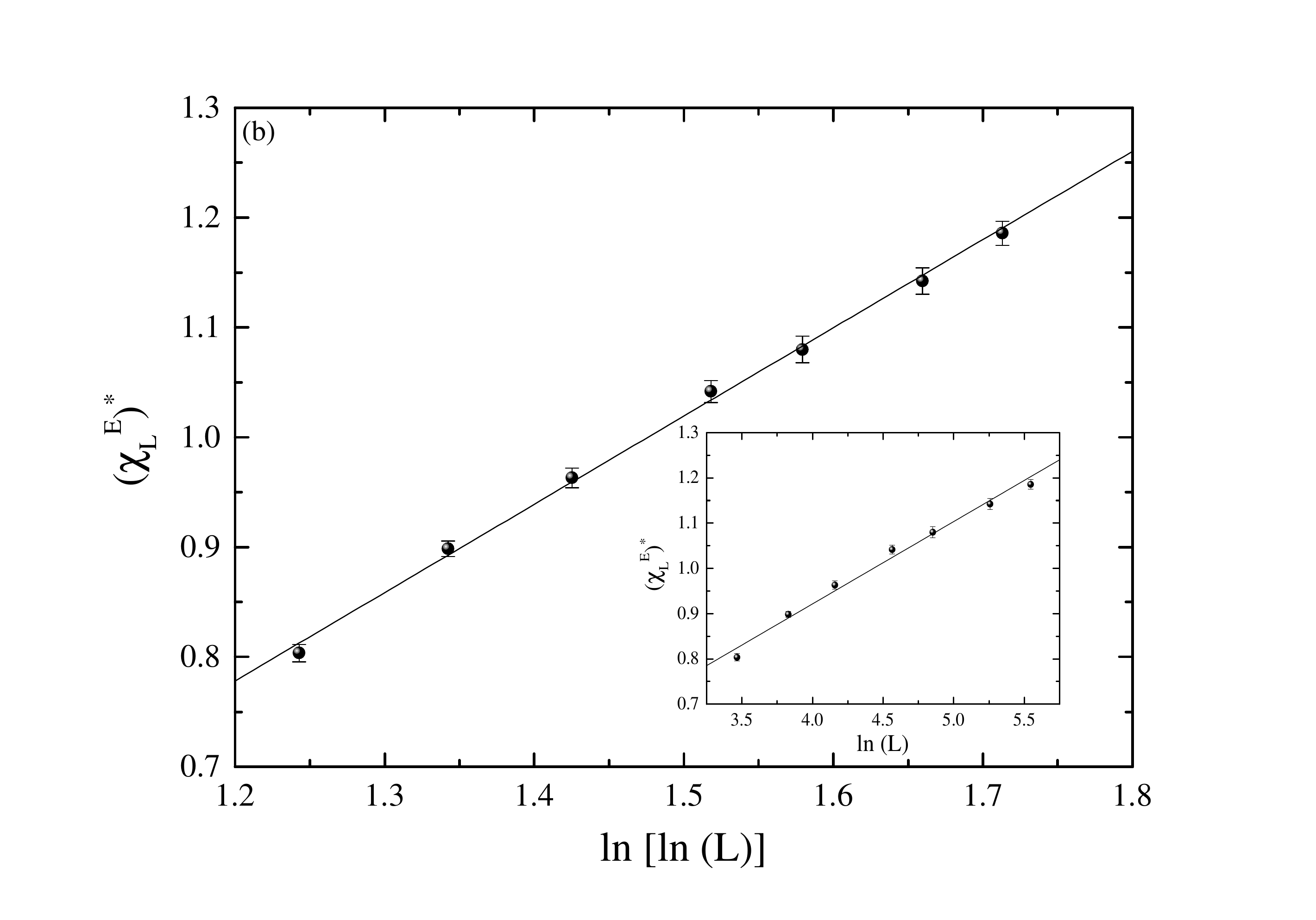}
\caption{\label{fig:Ising} Criticality in the kinetic random-bond
Ising model ($r = 1/7$): (a) Finite-size scaling behavior of the
maxima $(\chi_{L}^Q)^{\ast}$ in a log-log scale (main panel) and shift behavior of the corresponding pseudocritical half periods $t_{1/2}^{\ast}$ (inset). (b) Double (main panel) and simple (inset) logarithmic scaling behavior of the heat-capacity maxima $(\chi_{L}^E)^{\ast}$. In all cases lines are linear fittings.}
\end{figure}

Figure~\ref{fig:time_series_Ising} presents the time evolution of
the magnetization and Fig.~\ref{fig:series_Ising} the period
dependencies of the dynamic order parameter $Q$ of the kinetic
random-bond Ising model. Several comments are in order: For
rapidly varying fields, Fig.~\ref{fig:time_series_Ising}(a), the
magnetization does not have enough time to switch during a single
half period and remains nearly constant for many successive field
cycles, as also illustrated by the black line in
Fig.~\ref{fig:series_Ising}. On the other hand, for slowly varying
fields, Fig.~\ref{fig:time_series_Ising}(c), the magnetization
follows the field, switching every half period, so that $Q \approx 0$, as also shown by the blue line in
Fig.~\ref{fig:series_Ising}. In other words, whereas in the
dynamically disordered phase the ferromagnet is able to reverse
its magnetization before the field changes again, in the
dynamically ordered phase this is not possible and therefore the
time-dependent magnetization oscillates around a finite value. The
competition between the magnetic field and the metastable state is
captured by the half-period parameter $t_{1/2}$ (or by the
normalized parameter $\Theta = t_{1/2} /\tau$, with $\tau$ being
the metastable lifetime~\cite{Park}). Obviously, $t_{1/2}$ plays
the role of the temperature in the equilibrium system. Now, the
transition between the two regimes is characterized by strong
fluctuations in $Q$, see Fig.~\ref{fig:time_series_Ising}(b) and the evolution of the red line
in Fig.~\ref{fig:series_Ising}. This behavior is indicative of a
dynamic phase transition and occurs for values of the half period
close to the critical one $t_{1/2}^{\rm c}$ (otherwise when $\Theta \approx 1$). 
Of course, since the value $t_{1/2} = 76$ MCSS used for this illustration is
slightly above $t^{\rm c}_{1/2} = 74.7(3)$, see also
Fig.~\ref{fig:Ising}, the observed behavior includes as well some
nonvanishing finite-size effects.

Some additional spatial aspects of the transition scenarios
described above via the configurations of the local order
parameter $\{Q_{x}\}$ are shown in Fig.~\ref{fig:local_Ising}.
Below $t_{1/2}^{\rm c}$, see panel (a), the majority of spins
spend most of their time in the $+1$ state, \emph{i.e.}, in the
metastable phase during the first half period, and in the stable
equilibrium phase during the second half period, except for
equilibrium fluctuations. Thus most of the $Q_{x}\approx +1$ and
the system is now in the dynamically ordered phase. On the other
hand, when the period of the external field is selected to be
bigger than the relaxation time of the system, above $t_{1/2}^{\rm
c}$, see panel (c), the system follows the field in every half
period, with some phase lag, and $Q_{x}\approx 0$ at all sites
$x$. The system lies in the dynamically disordered phase. Near
$t_{1/2}^{\rm c}$ and the expected dynamic phase transition, there
are large clusters of both $Q_{x}\approx +1$ and $-1$ values,
within a sea of $Q_{x}\approx 0$, as shown in
Fig.~\ref{fig:local_Ising}(b).

Although the discussion above concentrated on the Ising case,
analogous description and relevant conclusions may be drawn also
for the dynamical ordering of the disordered-induced continuous
transition of the Blume-Capel model, as depicted in
Figs.~\ref{fig:time_series_BC} - \ref{fig:local_BC}. Note that in
this case the critical half-period of the system has been
estimated to be $t^{\rm c}_{1/2} = 83.6(4)$ (see also
Fig.~\ref{fig:BC} below). However, we should underline here that
for the case of the Blume-Capel model the value of the local order
parameter $\{Q_{x}\}$ does not distinguish between random
distributions of $\sigma_{x} = \pm 1$ and and regions of
$\sigma_{x} = 0$. To bring out this distinction, we present in
Fig.~\ref{fig:quad_BC} similar snapshots of the dynamic quadrupole
moment over a full cycle of the external field, $ O =
\frac{1}{2t_{1/2}}\oint q(t)dt$, where
$q(t)=\frac{1}{N}\sum_{x=1}^{N}\sigma_{x}^2$. In the spin-1
Blume-Capel model the density of the vacancies is controlled by
the crystal-field coupling $\Delta$ and, thus, the value of the
dynamic quadrupole moment changes depending on
$\Delta$~\cite{Vatansever_Fytas}. We point out that in
Fig.~\ref{fig:quad_BC}, except for the red $+1$ areas, the regions
enclosed by finite values demonstrate the role played by
the crystal-field coupling in the Blume-Capel model.

To further explore the nature of the dynamic phase transitions
encountered in the above disordered kinetic models we performed a
finite-size scaling analysis using the observables outlined in
Sec.~\ref{sec:observables}. Previous studies in the field
indicated that although scaling laws and finite-size scaling are
tools that have been designed for the study of equilibrium phase
transitions, they can be successfully applied as well to far from
equilibrium systems~\cite{Sides1,Sides2,Korniss,Buendia3,Park}.

\begin{figure}[t]
\includegraphics[width=10 cm]{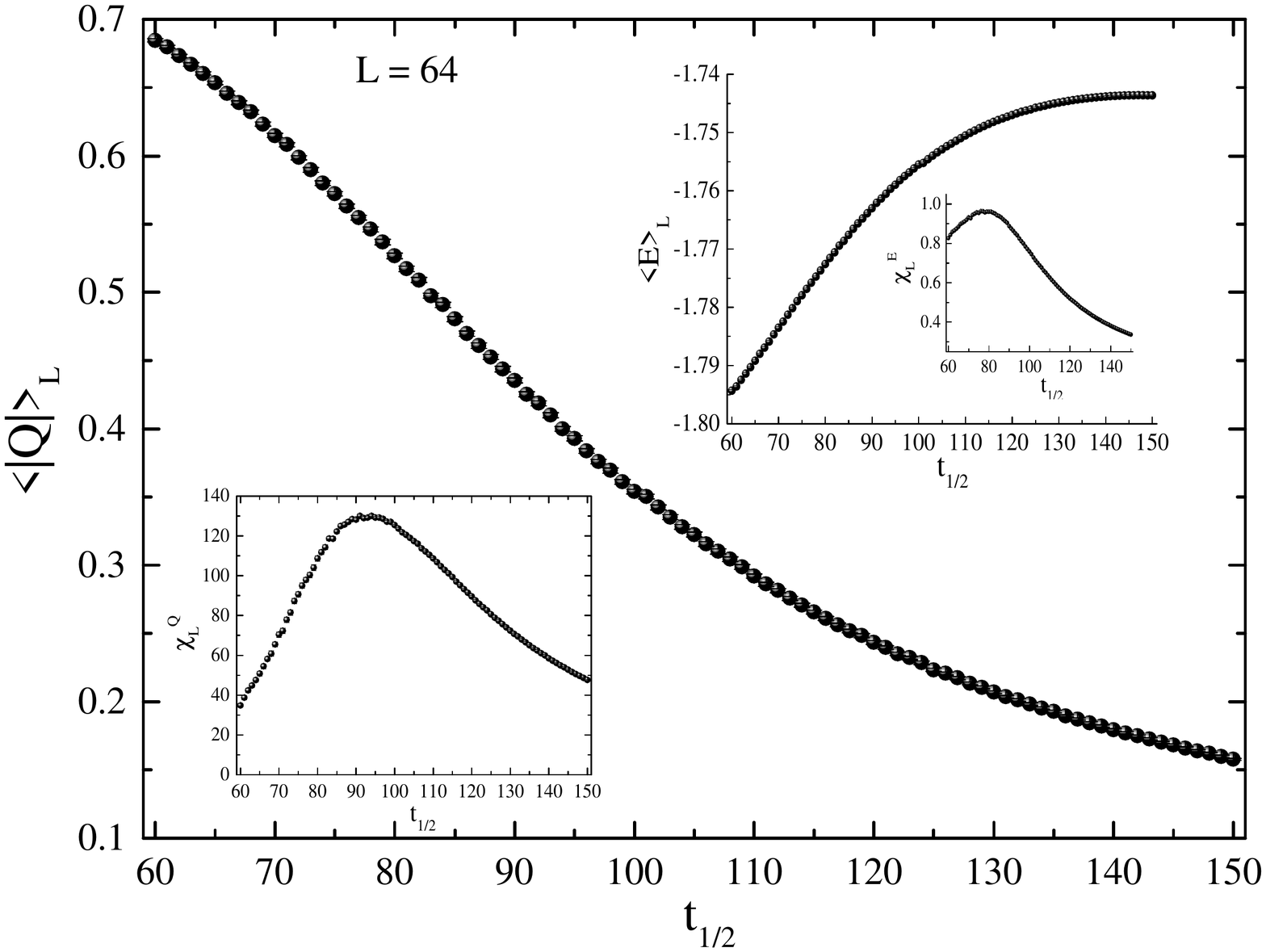}
\caption{\label{fig:plots} Half-period dependency of the dynamic
order parameter of the kinetic random-bond Ising model. The lower
inset illustrates the half-period dependency of the corresponding
dynamic susceptibility $\chi_{L}^{Q}$. The upper inset shows the
half-period dependency of the energy and the corresponding heat
capacity $\chi_{L}^{E}$. All results shown refer to a system size
$L = 64$ at the critical $t_{1/2}$-region.}
\end{figure}

As an illustrative example for the case of the kinetic random-bond
Ising model and for a system size of $L = 64$ we present in the
main panel of Fig.~\ref{fig:plots} the finite-size behavior of the
dynamic order parameter and in the lower inset the emerging
dynamic susceptibility [see Eq.(\ref{eq:7})]. The dynamic
order parameter goes from a finite value to zero values as the
half period increases showing a sharp change around the value of
the half period that can be mapped to the respective peak in the
plot of the dynamic susceptibility. The location of the maxima in
$\chi_{L}^{Q}$ may be used to define suitable pseudocritical half
periods, denoted hereafter as $t_{1/2}^{\ast}$. The corresponding
maxima may be analogously denoted as $(\chi^{Q}_{L})^{\ast}$.  
We also measured the energy and its scaled variance, the heat capacity $\chi_{L}^{E}$
[see Eq.(\ref{eq:8})]. The upper inset of
Fig.~\ref{fig:plots} shows the half-period dependency of the
energy of the same system and the relevant heat capacity. 
In this case the maxima may be denoted as
$(\chi_{L}^{E})^{\ast}$. 

We start the presentation of our finite-size scaling analysis with
Ising case. In the main panel of
Fig.~\ref{fig:Ising}(a) we present the size evolution of the peaks
of the dynamic susceptibility in a log-log scale. The solid line is a fit of the
form~\cite{Ferrenberg}
\begin{equation}
\label{eq:11} (\chi^{Q}_{L})^{\ast} \sim L^{\gamma/\nu},
\end{equation}
providing an estimate $1.75(1)$ for the magnetic exponent ratio
$\gamma/\nu$, in excellent agreement to the Ising value $7/4$. The
shift behavior of the corresponding peak locations
$t_{1/2}^{\ast}$ is plotted in the inset of
Fig.~\ref{fig:Ising}(a) as a function of $1/L$. The solid line
shows a fit of the usual shift form~\cite{Fisher,Privman,Binder92}
\begin{equation} \label{eq:12}
t_{1/2}^{\ast} = t_{1/2}^{\rm c} + bL^{-1/\nu},
\end{equation}
where $t_{1/2}^{\rm c}$ defines the critical half period of the
system and $\nu$ is the critical exponent of the correlation
length. The obtained values $t_{1/2}^{\rm c} = 74.7(3)$ and $\nu =
1.03(4)$ are listed also in the panel and, in particular, the
value of the critical exponent $\nu$ appears to be in very good
agreement to the value $\nu = 1$ of the 2D equilibrium Ising
model. This finding strongly supports the claim that the kinetic
Ising model under the presence of quenched bond randomness shares
the universality class of its corresponding equilibrium
counterpart. Ideally, we would also like to observe the
double logarithmic scaling behavior of the maxima of the heat
capacity $(\chi_{L}^{E})^{\ast}$. Indeed, as it is shown in
the main panel of Fig.~\ref{fig:Ising}(b), the data for the maxima of the heat capacity are adequately described by a fit of the form~
\begin{equation}
\label{eq:13} (\chi^{E}_{L})^{\ast} \sim \ln{[\ln{(L)}]},
\end{equation}
as predicted by Ref.~\cite{Dotsenko} for the random Ising
universality class. As a comparison, we plot the same data with respect to the simple logarithm of the system size in the corresponding inset. It is obvious that a fit $(\chi^{E}_{L})^{\ast} \sim \ln{(L)}$, as shown by the solid line, does not capture the full scaling behavior.

So, where do we stand at this point: We have shown that the
universality class of the dynamic phase transition encountered in
an Ising model under the presence of quenched bond disorder is
equivalent to that of its equilibrium counterpart with the
inclusion of logarithmic corrections in the scaling of the heat
capacity. We turn now our discussion to the dynamic phase
transition of the $\Delta = 1.975$ Blume-Capel model with
bond disorder. As mentioned previously in Sec.~\ref{sec:models}, 
only very recently the claims of universality violation in the
equilibrium random-bond Blume-Capel model have been dispelled and it
was shown that the induced under disorder continuous transition
belongs to the universality class of the random Ising model~\cite{Fytas17}. 
We therefore expect, or at least hope, that
the results presented in the current work will also be relevant to
this reignited problem, yet from a nonequilibrium perspective.

The scaling aspects of the dynamic phase
transition of the kinetic random-bond Blume-Capel model at $\Delta
= 1.975$ are shown in Fig.~\ref{fig:BC}, following fully the
presentation and analysis style of Fig.~\ref{fig:Ising} and excluding the data
for $L = 32$ that suffer from strong finite-size effects. In this
case an estimate $1.74(2)$ is obtained for the magnetic exponent
ratio $\gamma/\nu$, again compatible within errors to the Ising value
$7/4$. From the shift behavior of the corresponding pseudocritical half periods $t_{1/2}^{\ast}$ [inset of Fig.~\ref{fig:BC}(a)] the critical half-period and the correlation-length exponent are estimated to be $t_{1/2}^{\rm c} = 83.6(4)$ and $\nu = 1.05(7)$, respectively. Again the estimate of $\nu$ supports the
scenario presented above in Fig.~\ref{fig:Ising} for the
criticality in the dynamic phase transition of the random-bond
kinetic Ising model. Last but not least, in Fig.~\ref{fig:BC}(b)
the maxima of the heat capacity $(\chi_{L}^E)^{\ast}$ are plotted
versus $\ln{[\ln{(L)}]}$ (main panel) and $\ln{(L)}$ (inset) and as in the Ising case are much better described by the double logarithmic fit~(\ref{eq:13}).

\begin{figure}[t]
\includegraphics[width=10 cm]{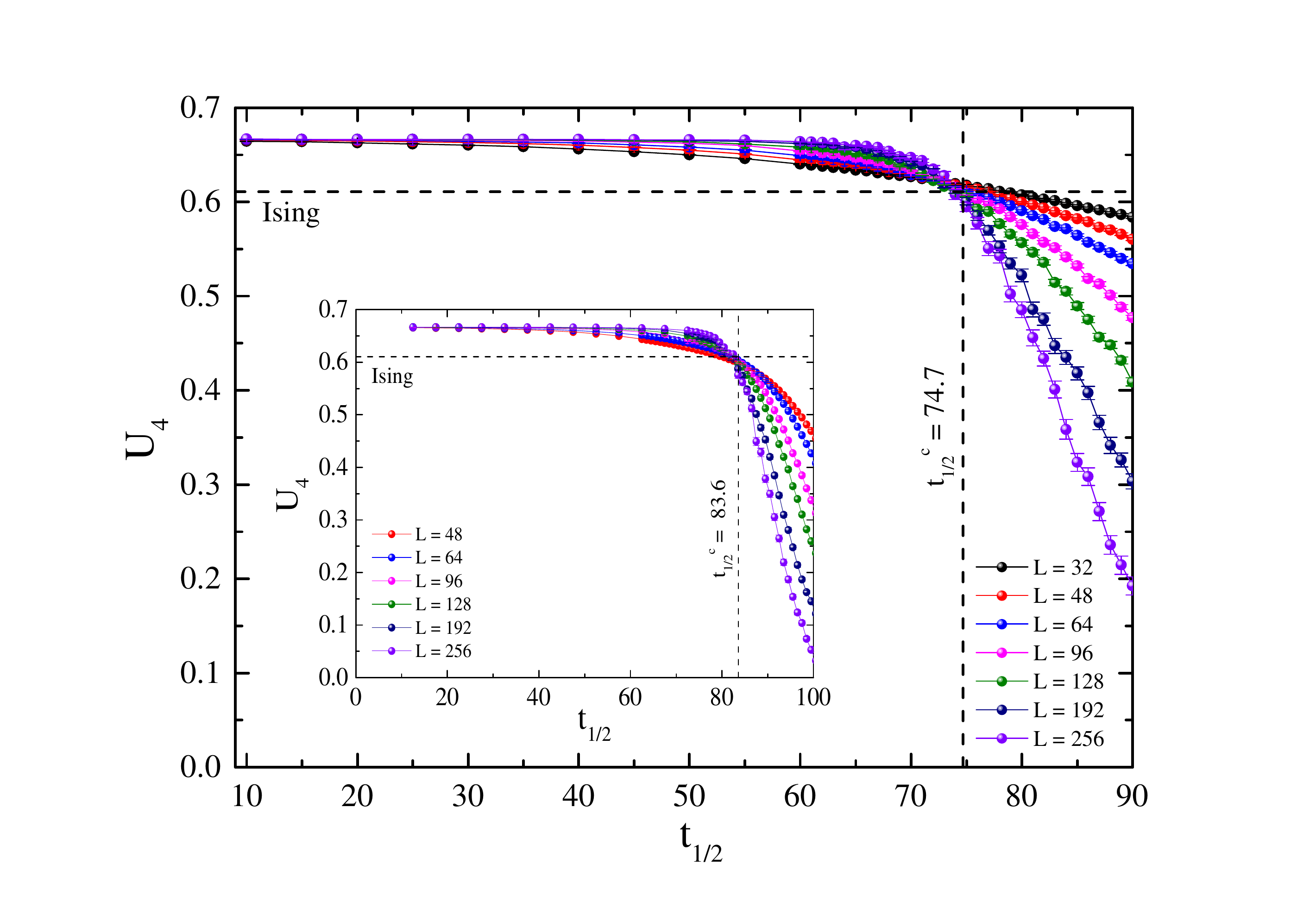}
\caption{\label{fig:Binder} Half-period dependency of the
fourth-order Binder cumulant $U_{L}$ of the kinetic random-bond
Ising (main panel) and Blume-Capel (inset) models for a wide range
of system sizes studied.}
\end{figure}

An alternative test of universality comes from the study of the 
fourth-order Binder cumulant $U_{L}$  defined in Eq.~(\ref{eq:9}) 
for the case of the dynamic order parameter. In
Fig.~\ref{fig:Binder} we present our numerical data of $U_{L}$ for
the kinetic random-bond Ising (main panel) and Blume-Capel (inset) models. 
In both panels the vertical dashed line marks the critical
half-period value of the system $t_{1/2}^{\rm c}$ and the horizontal
dotted line the universal value $U^{\ast} = 0.6106924(16)$
of the 2D equilibrium Ising model~\cite{Salas}. Certainly, the crossing point 
is expected to depend on the lattice size $L$ (as it also shown in the figure) 
and the term universal is valid for given lattice shapes, boundary conditions, and isotropic
interactions~\cite{Selke_2,Selke_3}. However, the data shown in Fig.~\ref{fig:Binder} support,
at least qualitatively, another instance of equilibrium Ising universality, since in both panels the 
crossing point is consistent to the value $0.6106924$. 
We should note here that Hasenbusch \emph{et al.}
presented very strong evidence that the critical Binder cumulant of the equilibrium 
2D randomly site-diluted Ising model maintains its pure-system value~\cite{Hasenbusch08}.  
In this respect, a dedicated study along the lines of Ref.~\cite{Hasenbusch08} for an 
accurate estimation of $U^{\ast}$ in the kinetic random-bond Ising and Blume-Capel models
would be welcome, but certainly goes beyond the scope of the current work.

\section{Conclusions}
\label{sec:conclusions}

In the present work we investigated the effect of quenched
disorder on the dynamic phase transition of kinetic spin models in
two dimensions. In particular, we considered the square-lattice
Ising and Blume-Capel models under a periodically oscillating
magnetic field, the latter at its low-temperature regime where the
pure equilibrium system exhibits a first-order phase transition. Using Monte
Carlo simulations and finite-size scaling techniques we have been
able to probe with good accuracy the values of the critical
exponent $\nu$ and the magnetic exponent ratio $\gamma/\nu$, both
of which were found to be compatible to those of the equilibrium
Ising ferromagnet. An additional study of the scaling behavior of
the heat-capacity revealed the double logarithmic divergence
expected for the universality class of the random Ising model. To
conclude, although universality is a cornerstone in the theory of
critical phenomena, it stands on a less solid foundation for the
case of nonequilibrium systems and for systems subject to quenched
disorder. In the current work we have studied two systems where
both of the above complications merge, yet arriving to the
simplest scenario. We hope that our work will stimulate further
research in the field of nonequilibrium critical phenomena at both
numerical and analytical directions.

\begin{acknowledgments}
The authors would like to thank P. A. Rikvold and W.
Selke for many useful comments on the manuscript. The numerical calculations reported in this paper were performed
at T\"{U}B\.{I}TAK ULAKBIM (Turkish agency), High Performance and
Grid Computing Center (TRUBA Resources).
\end{acknowledgments}

\newpage

\begin{figure}[t]
\centering
\includegraphics[width=10 cm]{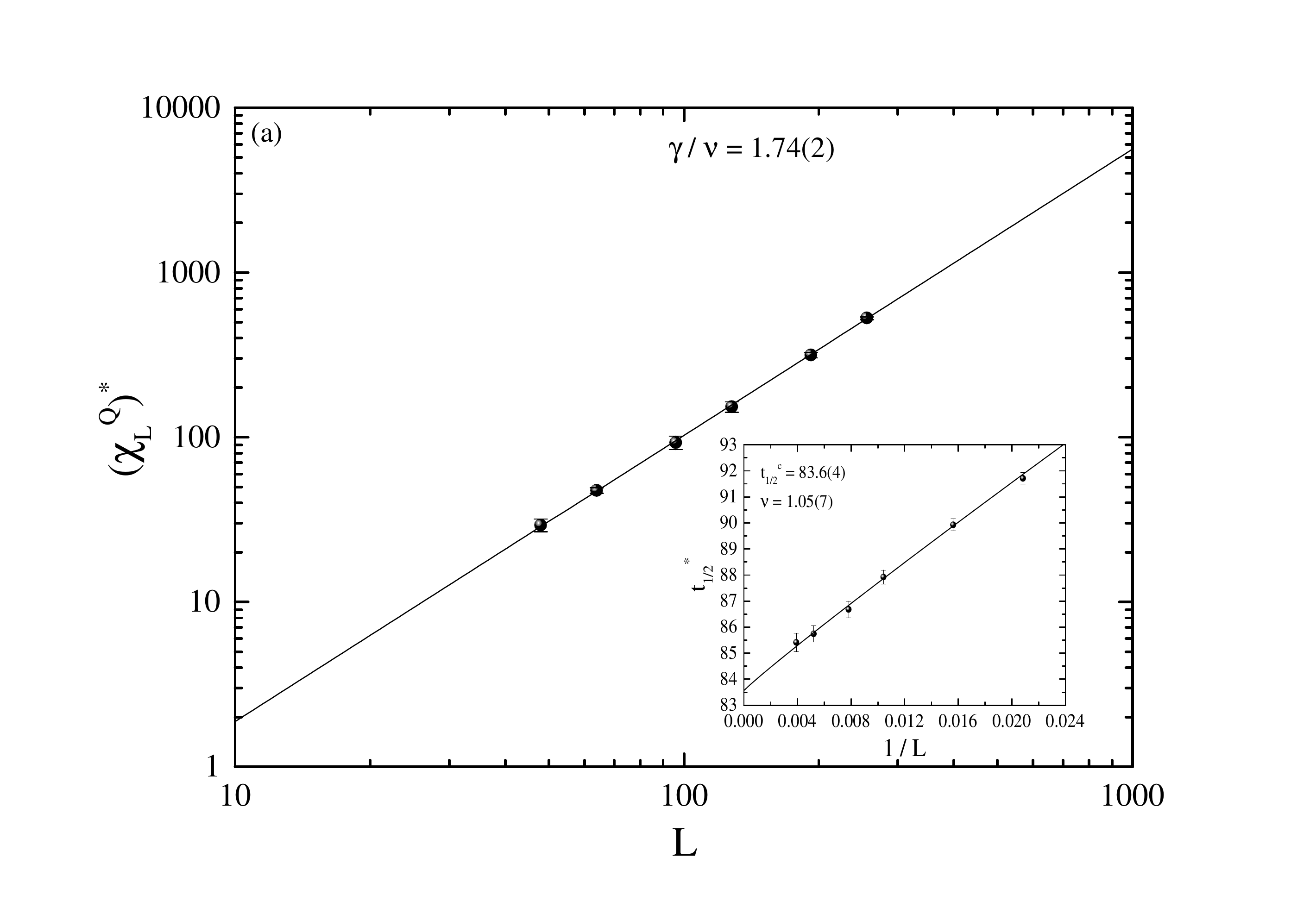}\\
\includegraphics[width=10 cm]{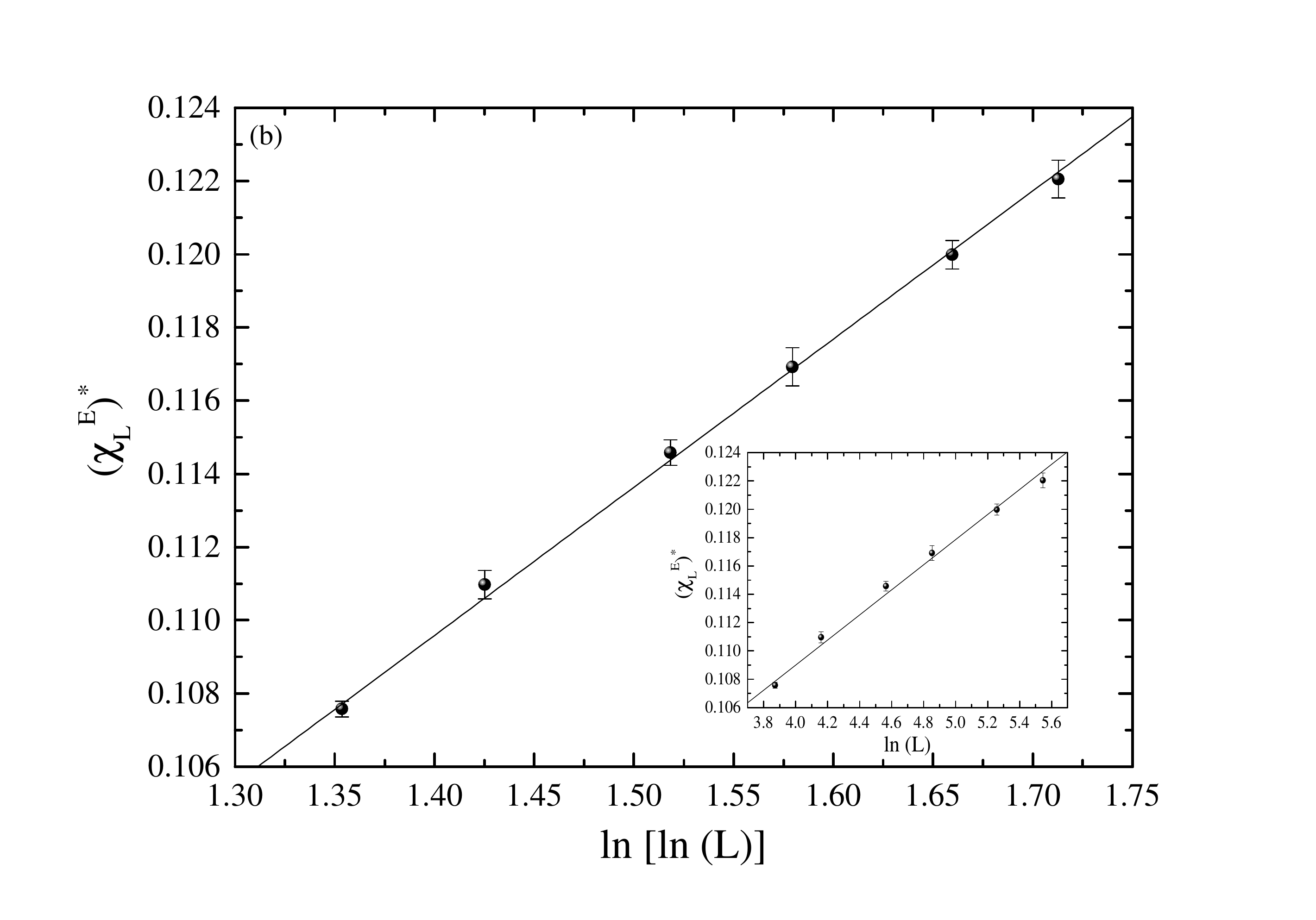}
\caption{\label{fig:BC} Criticality in the kinetic random-bond
Blume-Capel model ($r = 0.75/1.25$ ; $\Delta = 1.975$). The
description is analogous to that of Fig.~\ref{fig:Ising}.}
\end{figure}

\end{document}